\documentclass[linenumbers,a4paper,11pt]{article}
\pdfoutput=1 

\usepackage{jinstpub} 

\graphicspath{{figs/}}

\usepackage[colorinlistoftodos]{todonotes}

\newcommand{\cU}{\ensuremath{\mathcal{U}}}

\title{Kinematics reconstruction of the EAS-like events registered by the TUS detector}

\author[a]{Sergei~Sharakin,}
\author[b]{Oliver Isac Ruiz Hernandez,}
\affiliation[a]{Lomonosov Moscow State University, Skobeltsyn Institute
of Nuclear Physics,\\GSP-1, Leninskie Gory, Moscow, 119991, Russia}
\affiliation[b]{Benemérita Universidad Autónoma de Puebla, Facultad de Ciencias Físico Matemáticas,\\Avenida San Claudio y 18 Sur, C.P. 72570, Puebla, México}

\emailAdd{sharakin@mail.ru}

\abstract{
The Tracking Ultraviolet Set-up (TUS) is the world’s first orbital imaging detector of Ultra High Energy Cosmic Rays (UHECR) and it operated in 2016-2017 as part of the scientific equipment of the Lomonosov satellite. The TUS was developed and manufactured as a prototype of the larger project K-EUSO with the main purpose of testing the efficiency of the method for measuring the ultraviolet signal of extensive air shower (EAS) in the Earth's night atmosphere. 

Despite the low spatial resolution ($\sim5\times5$~km$^2$ at sea level), several events were recorded which are very similar to EAS as for the signal profile and kinematics.
Reconstruction of the parameters of such events is complicated by a short track length, an asymmetry of the image, and an uncertainty in the sensitivity distribution of the TUS channels.
An advanced method was developed for the determination of event kinematic parameters including its arrival direction. 

In the present article, this method is applied for the analysis of 6 EAS-like events recorded by the TUS detector. All events have an out of space arrival direction with zenith angles less than 40$^\circ$. Remarkably they were found to be over the land rather close to United States airports, which indicates a possible anthropogenic nature of the phenomenon. Detailed analysis revealed a correlation of the reconstructed tracks with direction to airport runways and Very High Frequency (VHF) omnidirectional range stations.

The method developed here for reliable reconstruction of kinematic parameters of the track-like events, registered in low spatial resolution, will be useful in future space missions, such as K-EUSO.

}

\keywords{ultra-high-energy cosmic rays, reconstruction,
cosmic ray detectors, TUS, orbital detector, Lomonosov satellite}

\begin{document}
\maketitle
\flushbottom

\section{The TUS detector and events with tracks}
\label{sec:Intro}

The TUS detector is a very sensitive telescope looking into the nadir direction from space, aimed at measuring extensive air showers (EAS) produced by ultra-high-energy cosmic rays in the atmosphere. This is the first attempt to perform the EAS measurement technique, proposed by John Linsley~\cite{Linsley1981}. The detector was launched on board the Lomonosov satellite on 28.04.2016 from Vostochny cosmodrome (Russian Federation) to the polar sun-synchronous orbit with height $R=470-500$~km and inclination 97.3$^\circ$.
The telescope consists of two main parts: the mirror-concentrator and the photodetector in its focal plane.  
The photodetector has 256 channels (``pixels'') 
disposed as a square matrix of 16 modules by 16 channels (pixel map). Each channel is represented by a photomultiplier tube (PMT) Hamamatsu R1463
with a square window (15~mm $\times$ 15~mm). An UV filter is placed in front of each PMT to limit the wavelength range of measurements (300-400~nm).

The effective focal distance of the mirror is $f=1.5$~m, so the angular resolution of one pixel is equal to 10~mrad, which corresponds to $\sim5$~km on the ground. The total field of view (FOV) of the detector is $9^\circ \times 9^\circ$ or about $80\times80$~km$^2$. 

The temporal resolution of the detector in the fastest mode of operation (EAS mode) is determined by a gate time unit, GTU = 0.8~$\mu$s. Each recorded event is a set of 256 digital signals represented by ADC counts (waveforms), with a duration of 256 GTUs ($\approx205$~$\mu$s).
More details about the detector and first results of measurements can be found in~\cite{SSR, tus-jcap-2017}.

The detector operated till late 2017 and registered almost 80,000 events in the EAS mode. Several preliminary EAS candidates have been selected in the TUS data earlier~\cite{2017arXiv170605369B, 2019EPJWC.21006006B}. 
The Lomonosov--UHECR/TLE Collaboration continues the TUS data analysis.
One of the main tasks of this analysis is to search for and study events in which the signals are similar to that expected from an EAS. In total, about 100 events with the typical signal profile shape and duration for an EAS have been selected to date.
However, only a few of them exhibit the EAS-like kinematics.

\subsection{EAS-like TUS events}

When analyzing the events measured by the TUS detector, several classes with common properties were identified, see for example,~\cite{tus-jcap-2017}. Some of the most mysterious events are the so-called EAS-like events, with a spatio-temporal pattern typical for the EAS fluorescence. 

For these events, it is possible to identify a large number of hit pixels (channels whose signal significantly exceeds the background) that line up on the pixel map along some direction (``thick'' track). The total signal of the hit pixels with a subtracted background (the light curve) has a characteristic fast increase with its subsequent slightly slower decline. The full duration of the signal is about 0.1~ms.

Six of the events registered by the TUS detector have these properties, see table~\ref{tab:TUSevents}.
Interestingly, all events were recorded over the United States and over land.

In \cite{Minnesota}, one of these events, TUS161003\footnote{Hereinafter TUS events are identified by indicating the registration date, so the first event in table~\ref{tab:TUSevents} was triggered on October 03, 2016. To distinguish the last two events recorded on the same day, suffixes a and b are used.}, was reconstructed to test the hypothesis of its EAS origin. However, an energy of this EAS candidate estimated by the amplitude of the luminescence is at least 1 ZeV according to simulation in the ESAF (EUSO Simulation and Analysis Framework) software, \cite{Berat2010}. 
Due to small total exposure (1200-1400 km$^2$ sr year) and a very steep spectrum of cosmic rays in this energy range, a chance the TUS to have registered such an extremely energetic cosmic particle is very low (order of $10^{-3}-10^{-5}$).
Some other possible explanations of the origin of this event were suggested, including anthropogenic sources.

\begin{table}[!ht]
\caption{List of EAS-like events and their location (geographical coordinates correspond to the center of the TUS FOV). The number of hit pixels is indicated in  the last column.}
\begin{center}\small
\begin{tabular}{|l|*{5}{c|}}
\hline
Event & Time (UTC) & Latitude & Longitude & Location  & \# hit pixels\\
\hline
TUS161003 & 05:48:59 & 44.08$^\circ$N & 92.71$^\circ$W & Minnesota (USA) & 10  \\
\hline
TUS161031 & 10:25:18 & 61.30$^{\circ}$N & 155.69$^{\circ}$W & Alaska (USA)  & 8\\
\hline
TUS170915 & 06:30:18 & 40.31$^{\circ}$N & 107.07$^{\circ}$W & Colorado (USA) & 12 \\
\hline
TUS171010 & 04:26:04 & 34.83$^{\circ}$N & 77.39$^{\circ}$W & North Carolina (USA) & 15 \\
\hline
TUS171029a & 06:39:09 & 35.27$^{\circ}$N & 110.78$^{\circ}$W & Arizona (USA)  & 8\\
\hline
TUS171029b & 11:13:26 & 65.90$^{\circ}$N & 168.07$^{\circ}$W & Alaska (USA) & 9 \\
\hline

\end{tabular}
\end{center}
\label{tab:TUSevents}
\end{table}

\begin{figure}
    \centering
    \includegraphics[width=0.75\textwidth]{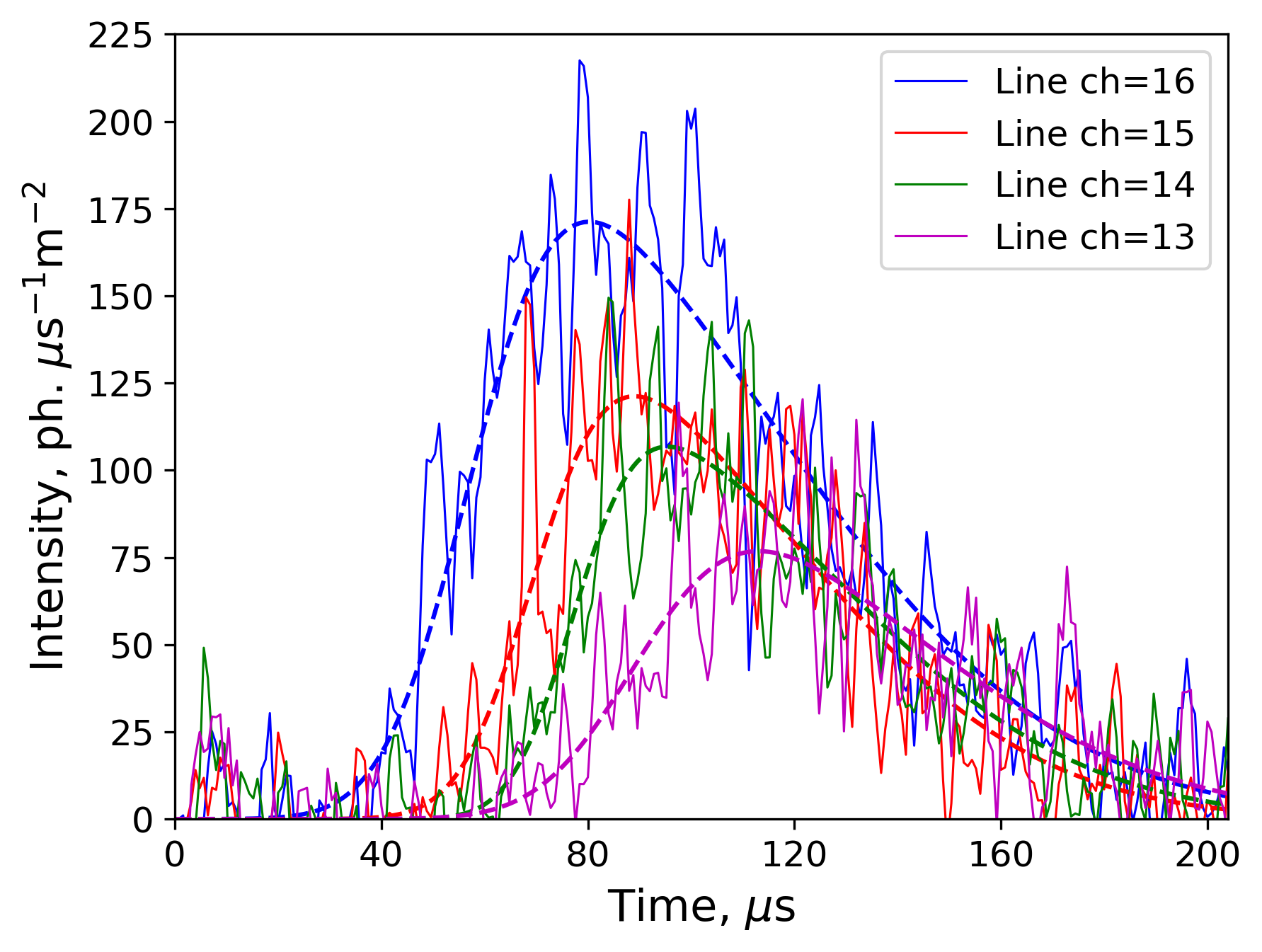}
 \caption{Peak movement for TUS170915 from line ch=16 (blue curve) to ch=13 (magenta). Dashed lines are results of the SkewGauss approximation.}
    \label{fig:TUS170915_motion}
\end{figure}

\begin{figure}
    \centering
    \includegraphics[width=0.95\textwidth]{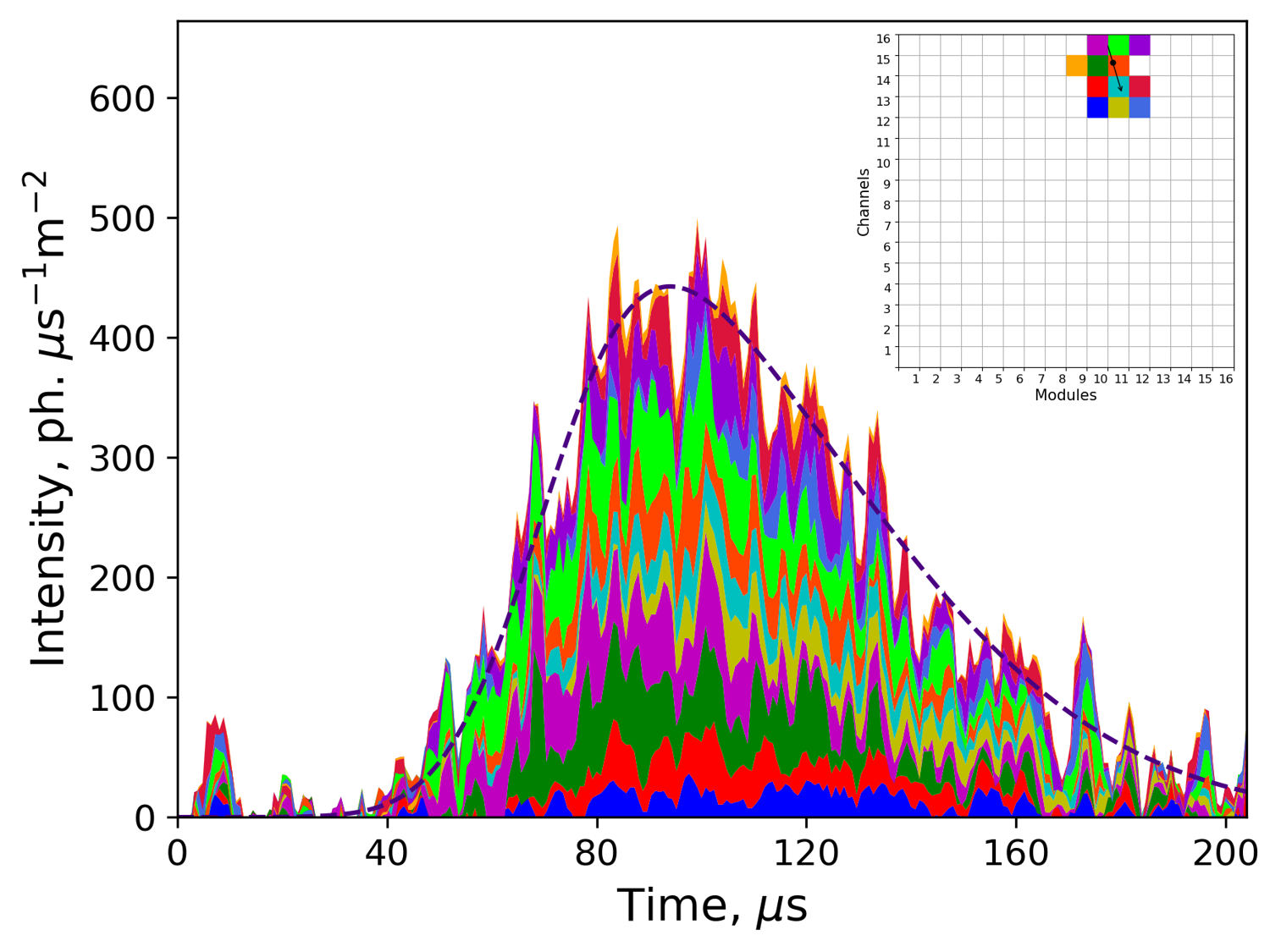}
    \caption{Light curve of the event TUS170915 and its SkewGauss approximation. Inset: Pixel map of selected pixels and reconstructed track.}
    \label{fig:sample_LC}
\end{figure}

In this paper, we summarize information about the EAS-like events by reconstructing their kinematics.
In addition to the features of the light curve and the image on the photodetector, these events are characterized by a noticeable time shifts of the signal peaks in the hit channels -- the motion of the image along the track.
Figure~\ref{fig:TUS170915_motion} shows the active signals of the TUS170915 (and their smooth approximations). Here we grouped 12 selected pixels (see insert in  figure~\ref{fig:sample_LC}) into 4 ``lines'' - directions perpendicular to the photodetector modules. 
The peak of the signals summed along the lines moves sequentially in time from the line ch = 16 to the line ch = 13.
Such movement is typical for a relativistic object with a luminescence region of the size of one channel or less on the FOV (``point-like'' source).

Hit pixels selection for each of these events was carried out by the amplitude of the moving average of the signal MA(8) (with a time window equals 8 GTUs). 
The value of the amplitude of the smoothed signal over its stationary value was studied at different thresholds, then the pre-selected channels were tested by location in space (neighboring pixels) and time (proximity to maximum).
The number of hit pixels is indicated in the last column of the table~\ref{tab:TUSevents}.

At the next stage of the analysis, the active signal extraction was made for each of the selected channels. To do this, the ADC counts was initially converted into the photon flux density at the entrance pupil of the detector, expressed as the number of photons per square meter per microsecond (for more details of this procedure, see ~\cite {Minnesota}). After that, the signal was approximated by a parametric function with an asymmetric shape of the waveform. The SkewGauss function (the density function of the Skew-Normal distribution \cite{Skew_gauss}) was chosen for this, and an additional constant was added to evaluate a background (Base Level of the signal).

Active signal extraction allows us to reconstruct the light curve of the event. To quantify the latter, the sum of the all active signals was approximated by the SkewGauss function again. For example, figure~\ref{fig:sample_LC} shows the result of such a parameterization over 12 selected channels of the TUS170915.
The signal value at the maximum (amplitude of the light curve) was $A_\mathrm{m}=442$~ph/($\mu$s\,m$^2$), which is approximately two times more than that of the event TUS161003. In this paper, we will characterize the luminescence duration by the full duration at half maximum (FDHM), and its asymmetry by the ratio of the signal decay time to its rising  time (taken at the half-amplitude level). For the event TUS170915, FDHM = 72~$\mu$s and Asymmetry =1.69.

The reconstruction of the light curves of the  others EAS-like TUS events (the results are given in section~\ref {sec:TUSreco}) gives the FDHM 50--70 $\mu$s, and the asymmetry varying from 1.5 to 2.2. Estimates of the amplitude of the signals also have significant variations.

As mentioned above, the time shifts of the peak of the pixel's signal can be interpreted as a movement of a relativistic luminous compact object (in comparison to the resolution of the detector). To determine the origin of the EAS-like events, it is important to reconstruct the arrival direction of these objects.

\subsection{Arrival direction}

The optical system of the detector projects the movement of the source object onto the image's movement along the track on a focal plane. In the case of ideal (aberration-free) optics, the speed of the point-like image $U$ is related to the apparent angular velocity of the light source $\omega$ by means of the focal distance $f$,  $U = \omega \cdot f$.

If the point-like light source moves in the atmosphere at a constant speed along a rectilinear path, its image will also move along the rectilinear track on the focal plane. 
More precisely, the track is formed by non-aberration points (we will call them ``track points''), the structure of the whole image depends on the point spread function (PSF) of the optics.
The direction of motion of the luminous object (arrival direction) can be determined by calculating the components of the speed of the non-aberration point along the track.

In the local coordinate system, the direction of arrival $\vec\Omega$ can be set using azimuth $\phi$ and zenith $\theta$ angles:
$\vec\Omega = \{ \sin\theta \cos\phi, \sin\theta \sin\phi, \cos\theta \}$ ($\vec\Omega$ directs opposite to velocity of the source). 

If we denote by $\chi$ the angle between $\vec\Omega$ and line of sight $\vec s$, the
angular motion of the source can be expressed through the speed of the source~$v$ and the distance from it to the detector~$R$:
\begin{itemize}
    \item \textit{Non-relativistic sources} ($v\ll c$): $\omega R = v\sin\chi$
\item \textit{Relativistic sources} ($v\sim c$):
$\omega R = v\sin\chi/(1+\frac{v}{c}\cos\chi)$
\item 
\textit{Ultra-relativistic sources} ($v\approx c$): $\omega R = c\tan(\chi/2)$
\end{itemize}

When observed in nadir, $\chi\approx\theta$. For off-axis events,
$\cos\chi = \cos\gamma \cos\theta - \cos(\delta\phi) \sin\gamma \sin\theta$, where $\gamma$ is the field angle and $\delta\phi$ is the angle between $\vec\Omega$ and $\vec s$ projected onto the ground. 

For a small $\gamma$, the difference between $\chi$ and $\theta$ can be estimated as a small correction. Thus for ultra-relativistic sources in the local geographical coordinate system, we finally find:
\begin{equation}\label{eq:Theta}
\phi = \Phi_\mathrm{det} - \mathrm{atan}(U_y/U_x), \quad \theta=2\,\mathrm{atan}(U/U_0)+\gamma \cos(\delta\phi)
\end{equation}
with $U_0=fc/R$ and $\Phi_\mathrm{det}$ is the azimuth  direction of the detector movement. For the TUS detector, $U_0=0.9$~mm/$\mu$s at $R=500$~km (or approximately 1 pixel per 20 GTUs) and the heading angle $\Phi_\mathrm{det}\sim -160^\circ$ for all six EAS-like events.

It follows from~(\ref{eq:Theta}) that the  movement along the track is non-uniform, however as the only brightest part of the track could be detected, the variation in speed can be ignored. So to estimate the zenith angle, we can substitute into equation (\ref{eq:Theta}) for parameters $R$, $\gamma$ and $\delta\phi$ their values at
maximum point, so
$\tan\gamma\approx \sqrt{X_\mathrm{m}^2+Y_\mathrm{m}^2}/f$, $\tan\delta\phi\approx Y_\mathrm{m}/X_\mathrm{m}$, where $(X_\mathrm{m},Y_\mathrm{m})$ are coordinates of the maximum point.

An important advantage of the orbital detection is its weak dependence to the height of the source. Indeed, to calculate the angle with 5\% accuracy, it is sufficient to know $R$ with the same accuracy, i.e. at accuracy 10~km for the  satellite's orbit $\sim500$~km. At the same time due to the exponential decrease in air density with altitude, the glow is localized in the lower layers of the atmosphere. 
For example, the height of the EAS maximum with an energy $\sim10^{20}$~eV does not exceed 15 km up to $\theta=80^\circ$.

The application of classical track reconstruction methods in this case is complicated by several reasons: a small number of track pixels, a large size and asymmetry of the PSF, and a significant uncertainty in the sensitivity of channels and PSF. 
For example, one can track the movement of the image centroid by approximating the hit pixel signals at each GTU with a suitable parameterized distribution. However, since the error in the localization of the centroid is comparable to the size of the entire image, for some events the displacement cannot be detected, and in some cases, on the contrary, the instantaneous velocity varies greatly from GTU to GTU.
Therefore, in ~\cite{Minnesota}, the Linear Track Algorithm heuristic (LTA) was proposed, which can be used to reconstruct track parameters in such an uncertainty situation.

In the following, section~\ref{sec:LTA} provides an analysis of the accuracy of the LTA, and section~\ref{sec:TUSreco} presents the results of its application to the EAS-like events registered by the TUS.
Finally, in section~\ref{sec:Disc}, localization of the reconstructed tracks allows us to hypothesize the origin of these events.

\section{LTA reconstruction scheme}
\label{sec:LTA}
The problem of reconstructing the image of the EAS fluorescence as a linear track on the focal surface of an orbital detector was studied in detail in a number of works of the JEM-EUSO collaboration, \cite{Berat2010}, \cite{BiktemerovaGuzmanMernik2013}. Two requirements are necessary to use the methods proposed in these papers: 1) matching of the optics and the photodetector and 2) a large number of hit pixels in the image. Here by matching, we mean that the characteristic size of the (instantaneous) image is approximately equal to the size of the photo sensor, which leads to a fairly accurate localization of the track point. A large number of pixels is required for small statistical errors of resulting estimates (for this reason, the accuracy of the arrival direction reconstruction decreases significantly with decreasing zenith angle). However, both of these requirements are not met in the case of the TUS detector. The size of the PSF is rather big so the instant image is distributed over 4-6 neighboring pixels. The track length does not significantly exceed its ``thickness'' (as we will see later, this is due to the fact that the recorded events have zenith angles of $30^\circ $ and less).

In an alternative approach, reconstruction of the EAS-like event can be achieved in a probabilistic model by Bayesian inference \cite{sivia2006data} of the posterior distribution to unknown track $\{ T_\mathrm{m}, X_\mathrm{m}, Y_\mathrm{m}, U_x, U_y \}$ and light curve parameters \{$A_\mathrm{m}$, FDHM, Asymmetry\}. The following information on the signals in the hit pixels is used in this case: the coordinates of the centers of these pixels $\{X_i, Y_i \} $, $ i = 1, ..., N $, and the values of the active signal in them $\{A_i( T_k)\} $ ($ N $ is the number of selected pixels, $k$ is the GTU index).
The active signal $A_i$ itself must be obtained by subtracting the background and subsequent normalization to the channel sensitivity.
The sensitivity coefficients and PSF are used as a priori information in such a probabilistic approach.
Unfortunately, this information is known with large uncertainty. The sensitivity of individual channels changed a lot during the operation of the detector in orbit (for more details, see \cite{Minnesota}) and it has to be estimated indirectly during experiment by stationary waveforms analyses. Lastly, the PSF of the TUS mirror varies greatly throughout the photodetector FOV, both in size and shape.

\subsection{LTA and its hyperparameters}

In the situation of such a strong uncertainty of a priori information, it is more efficient to develop a heuristic approach for track reconstruction. The idea of this approach, first described in \cite{TkachevICRC2017} and then modified in \cite{Minnesota}, is to make an approximation of the time dependence of the track points coordinates, $X(T) = X_\mathrm{m} + U_x (T - T_\mathrm{m})$, according to the distribution of the coordinates $ X_i$ of the centers of hit pixels: $T_\mathrm {m}$ is estimated by the light curve fit, the parameters $U_x$, $X_\mathrm {m}$ are selected by minimizing the weighted sum of the squared deviations (and similarly for $Y (T)$). To reduce the approximation roughness associated with pixelization, the corresponding signal values $A_i(T_k)$ are used as the weights of each ``point''.
More precisely, the algorithm includes the hyperparameter $n_\mathrm{exp}$ that controls a dependence of the weight $w$ on the signal, $w_{ik}\sim [\tilde A_i(T_k)]^{n_\mathrm{exp}}$, where $\tilde A_i(T)$ is (smoothed) value of the signal at time $T$.

Another important feature of this method, called Linear Track Algorithm (LTA), is the selection of a suitable time ``activity window'' for each signal. 
The very wide window leads to a systematic underestimation of $U_x$ while a small window corresponds to big variation of the results (see below). In this paper, it is proposed to control the size of the activity window by the second hyperparameter of the algorithm - the activity threshold $q_\mathrm{act}$. Thus, in LTA, the following sum is minimized

\begin{equation}
    \sum_{i=1}^N\sum_{k_1(i)}^{k_2(i)}  [\tilde A_i(T_k)]^{n_\mathrm{exp}}\cdot [X_\mathrm{m} + U_x(T_k -T_\mathrm{m}) - X_i]^2,
\end{equation}
by parameters $U_x$ and $X_\mathrm{m}$. The boundaries of the activity window $k_1(i)$, $k_2(i)$ are determined so that $\tilde A_i(T)$ exceeds the threshold $q_\mathrm{act}$ over the entire (maximum) interval $ [T_{k_1(i)}, T_{k_2(i)}]$.

Thus, the operation of the LTA is controlled by two hyperparameters, $n_\mathrm{exp}$ and $q_\mathrm{act}$, the dependence of the reconstruction results and the choice of their optimal values was carried out on a simulated events data base.

\subsection{Simulated events reconstruction}

To set the hyperparameters of the algorithm and to evaluate the accuracy of its predictions (reconstruction accuracy), we use a data base of simulated events with characteristics which are similar to the TUS EAS-like events. This determines the choice of the range of zenith angles, the duration and asymmetry of the light curve and the characteristic signal-to-noise ratio (associated with the number of triggered pixels).
For convenience, the track and light curve parametrization of the simulated events were performed directly on focal plane of the photodetector.
To generate a reference data base (RDB) with 1000 simulated events, we choose the following parameters distribution\footnote{$x\sim\cU[a,b]$ means $x$ uniformly distributed on the interval $[a,b]$.}:
\begin{itemize}
    \item Track: velocity projections $U_x$, $U_y$ corresponding to $\phi_0\sim \cU[-180^\circ,+180^\circ]$, $\theta_0\sim \cU[20^\circ,40^\circ]$.
    \item Light Curve: Skew-Gauss function with 
    uniformly distributed scale and shape parameters corresponding to FDHM$\in[52, 80]$ $\mu$s and Asymmetry$\in[1.5, 2.3]$.
\end{itemize}

The pixel distribution of the signal was performed using a 2D isotropic Gaussian PSF. The characteristic size of the PSF was selected to correspond to the  control optical tests~\cite{ICRC2013TUS}.
Measurements showed that the diameter of the circle of confusion with 70\% of full energy of the image $d_{70}$ varies from 15~mm in the center of the photodetector to 23-27~mm at the edge. Since all the TUS events covered in this article are localized close to the edge of the FOV, for the RDB we selected $d_{70}\sim \cU[22.5, 25.5]$ mm (which corresponds to the range 1.5-1.7 in pixel size units).
In order to take into account the uncertainty in the channel response, each sensitivity coefficient was varied due to a Gaussian distribution with a 20\% standard deviation.

The choice of the signal amplitude, background level and the pixel selection criteria were carried out, so that the number of such channels and the signal-to-noise ratio were comparable with those obtained for the TUS events. 
Accordingly, we fixed an amplitude $A_\mathrm{m}=100$~a.u. (arbitrary units) and the noise $\sigma_\mathrm{noise}=5$~a.u., so the pixel selection by their amplitude (both as MA(8) and as Gauss fit) at the level $q_\mathrm{sel}=4$~a.u. gives the number of the hit pixels averaged on RDB $N_\mathrm{sel}=7.0\pm1.1$.

The Signal-to-Noise ratio for the selected pixels is presented in table~\ref{tab:S2N_new}, for three variants of the 'signal': the amplitude of the Gauss fit, the signal integrated in 8 GTUs, and the signal inside FDHM interval (as a ratio to the corresponding standard deviation of the background).

\begin{table}[!ht]
\caption{Signal to Noise ratio (SNR) in first 7 selected pixels (sorted by amplitude) averaged on RDB for $q_\mathrm{sel}=4, \sigma_\mathrm{noise}=5$, the three variants correspond to fit amplitude, MA(8) selection and FDHM signal (see text).}
\begin{center}\small
\begin{tabular}{|l|*{7}{c|}}
\hline
SNR: & \#1 & \#2 & \#3 & \#4 & \#5 & \#6 & \#7  \\
\hline
$I_\mathrm{p}/\sigma_\mathrm{noise}$ & 7.4 & 5.2 & 3.7 & 2.8 & 1.8 & 1.3 & 0.8  \\
\hline
$S_8/(\sigma_\mathrm{noise}\sqrt{8})$ & 22.8 & 16.3 & 12.3 & 9.9 & 6.7 & 5.0 & 3.3 \\
\hline
$S/(\sigma_\mathrm{noise}\sqrt{\Delta T})$ & 46.1 & 32.5 & 23.7 & 18.1 & 11.1 & 7.8 & 4.9 \\
\hline
\end{tabular}
\end{center}
\label{tab:S2N_new}
\end{table}

The result of the arrival direction reconstruction of an individual simulated event can be presented as errors -- the differences between the reconstructed angles and their true values, $\Delta\phi\equiv\phi_\mathrm{rec}-\phi_0$ and $\Delta\theta\equiv\theta_\mathrm{rec}-\theta_0$.

Reconstruction of the entire RDB allows one to evaluate the  statistical accuracy (or reliability) of the method. For example, at fixed values of the LTA hyperparameters it can be presented as a histogram of errors.
The two main characteristics of the error distribution are the bias and width of the error histogram, which can be taken as the mean values of $\Delta\phi$ or $\Delta\theta$ and their standard deviations, respectively.

\begin{figure}[t]
\centering
\includegraphics[width=0.48\textwidth]{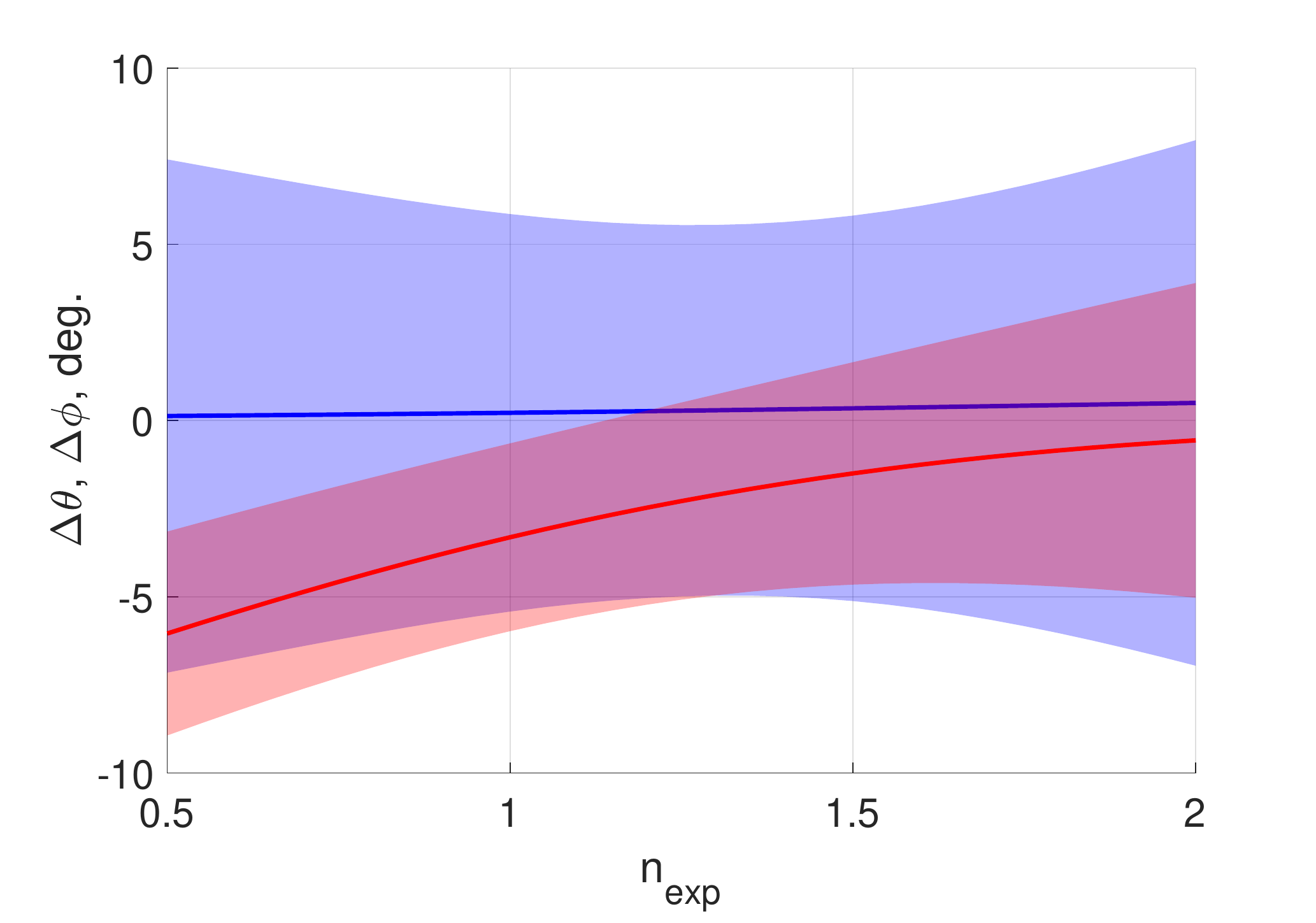}
\includegraphics[width=0.48\textwidth]{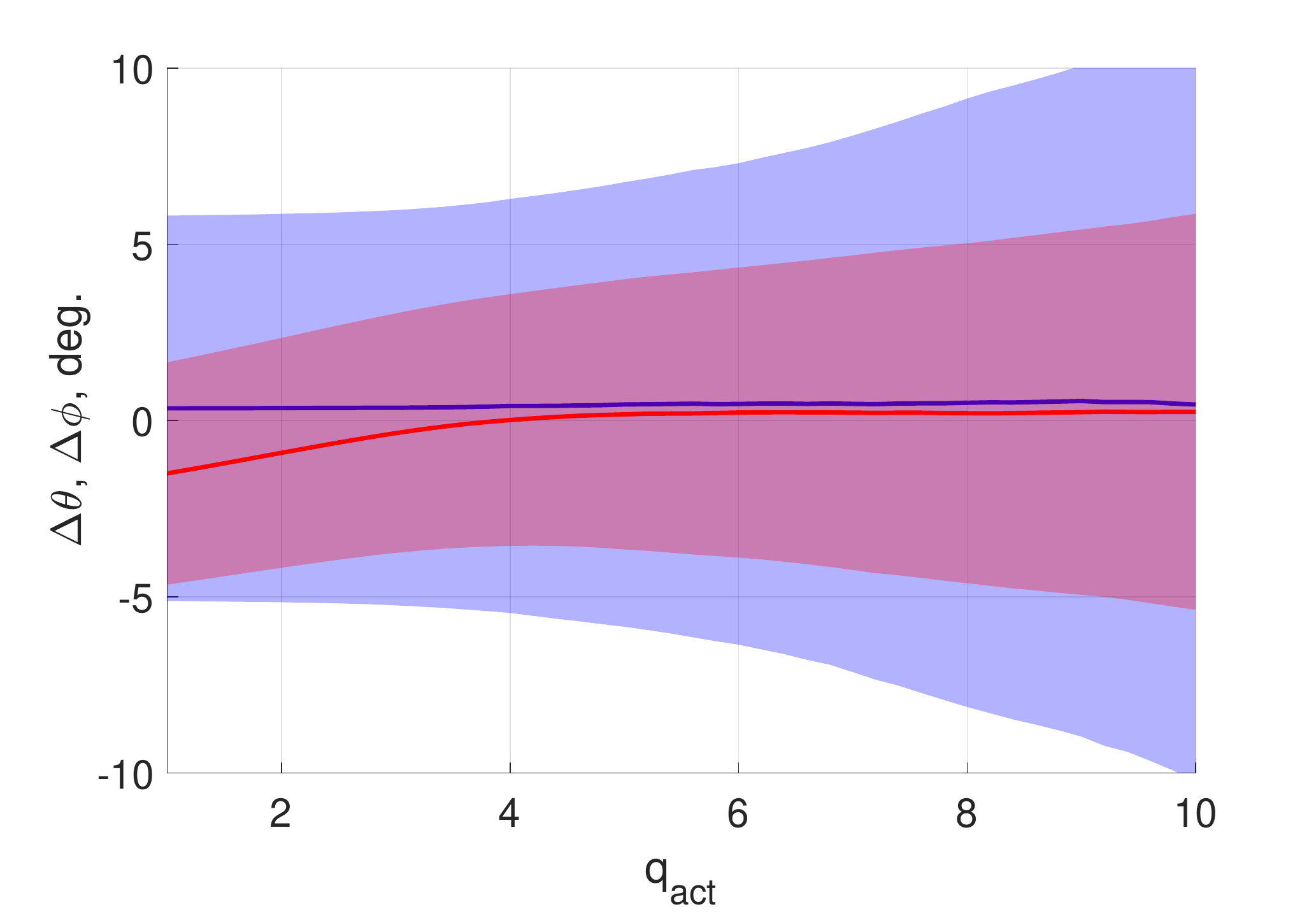}
\caption{LTA accuracy as a function of exponent~$n_\mathrm{exp}$  (left, $q_\mathrm{act}=5$~a.u.) and as a function of activity threshold~$q_\mathrm{act}$ (right, $n_\mathrm{exp}=1.5$). Blue colors for $\phi$-errors, red ones for $\theta$-errors.}
\label{fig:RDB_EGs}
\end{figure}

To study how the accuracy of the LTA depends on the hyperparameter setting, error curves were plotted for RDB -- the bias and width dependencies on $n_\mathrm{exp}$ and $q_\mathrm{act}$ for both angles, figure~\ref{fig:RDB_EGs} (the lines correspond to the mean values, the filled areas are $\pm1$ standard deviation, blue color for $\Delta\phi$, red one for $\Delta\theta$).
The left panel of the figure shows the error curves for the exponent parameter, which suggests an optimal choice for $n_\mathrm{exp}$. The position of the minimum of the errors, calculated as RMS $= \sqrt{\mathrm{bias}^2+\mathrm{width}^2}$, depends on the sensitivity variation, but is always close to $n_\mathrm{exp}\approx1.5$. In further work, we fix the hyperparameter of the exponent at this value.

The optimal values of the activity threshold $q_\mathrm{act}$ for reconstructing the azimuth angle lie in the region of small values, see the right panel of figure~\ref{fig:RDB_EGs}, while for the zenith angle this region has a significant bias (underestimating the speed along the track). On the other hand, increasing $q_\mathrm{act}$, the number of approximation points decreases, which leads to a wider error distribution.

\subsection{Event-by-event reconstruction}

\begin{figure}
\centering
\includegraphics[width=0.48\textwidth]{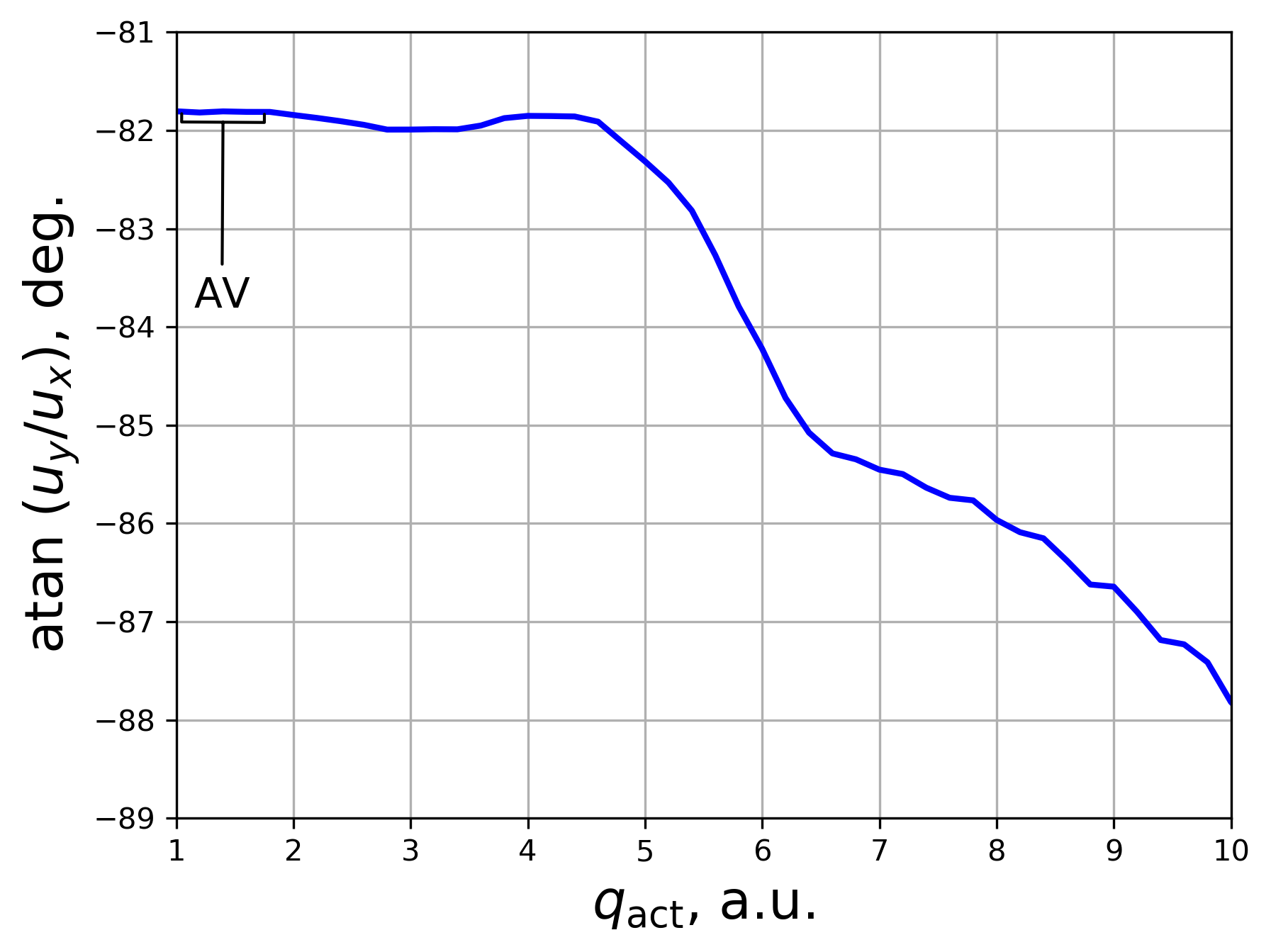}
\includegraphics[width=0.48\textwidth]{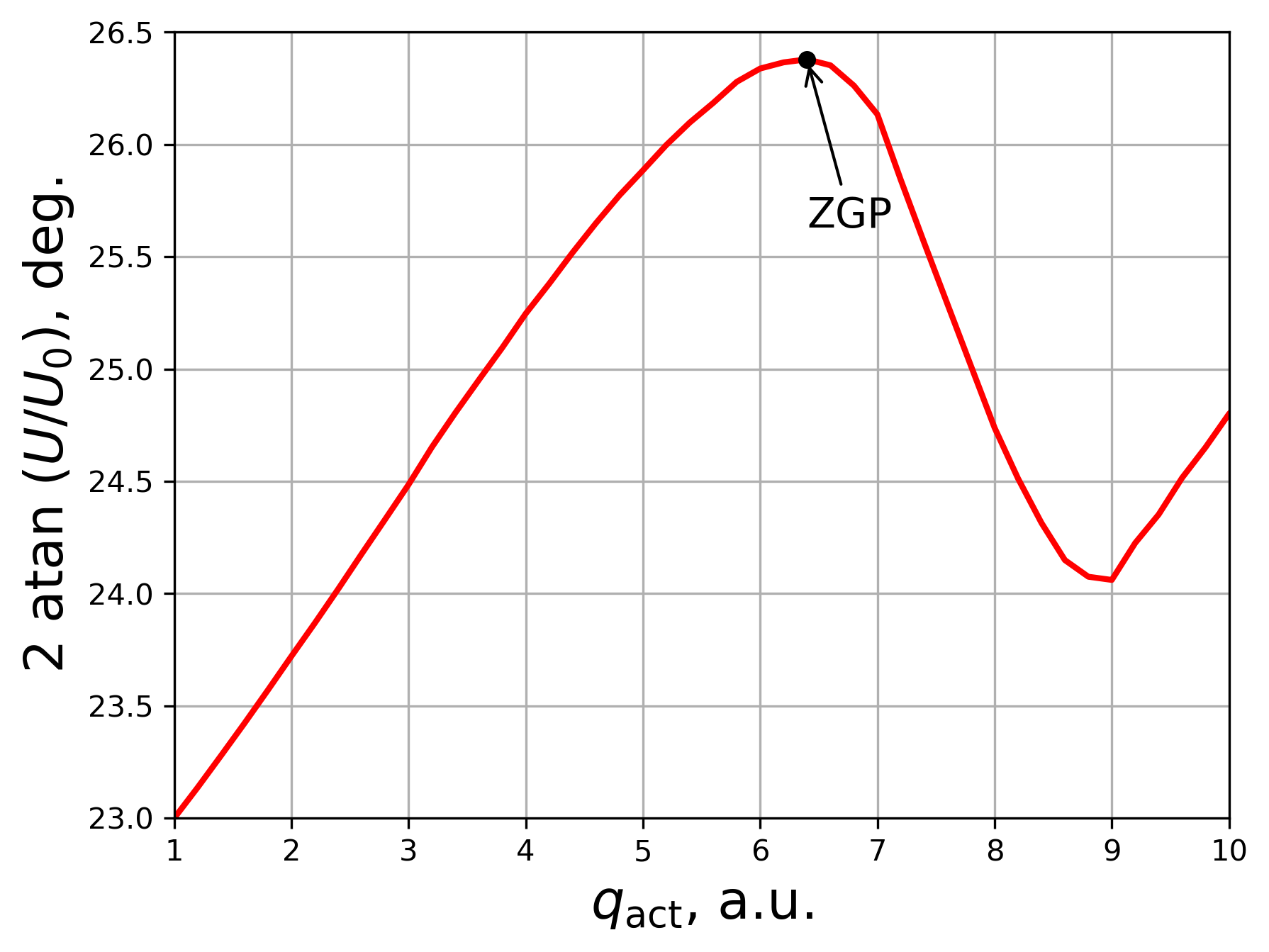}
\caption{Typical behaviour of a reconstruction curve for simulated event: $\phi$-curve with AV region at small threshold (left panel) and $\theta$-curve with ZGP (right panel)}
\label{fig:EG_typical}
\end{figure}

The optimal activity threshold cannot be fixed on one specific value - it will change when varying, for example, the signal-to-noise ratio or selection procedure. Therefore the final choice of the $q_\mathrm{act}$ should be carried out in event-by-event basis and rely on the behavior of reconstruction curves, i.e. the reconstructed angle dependence on the activity threshold.
The typical behavior for the RDB events reconstruction curves are shown in figure~\ref{fig:EG_typical}.

The shape of the $\phi$-curve has a region at small $q_\mathrm{act}$ with an asymptotic behaviour of the curve, the region with small $\phi_\mathrm{rec}$ variations. In the following, we called the choice of this asymptotic value for azimuth angle as AV-prescription. On the other hand, the $\theta$ reconstruction curve can have different shape with several local maxima. The first maximum (with the smallest $q_\mathrm{act}$) we called zero gradient point (ZGP), and use this ZGP-prescription to choose the $\theta_\mathrm{rec}$.

\begin{figure}
\centering
\includegraphics[width=0.48\textwidth]{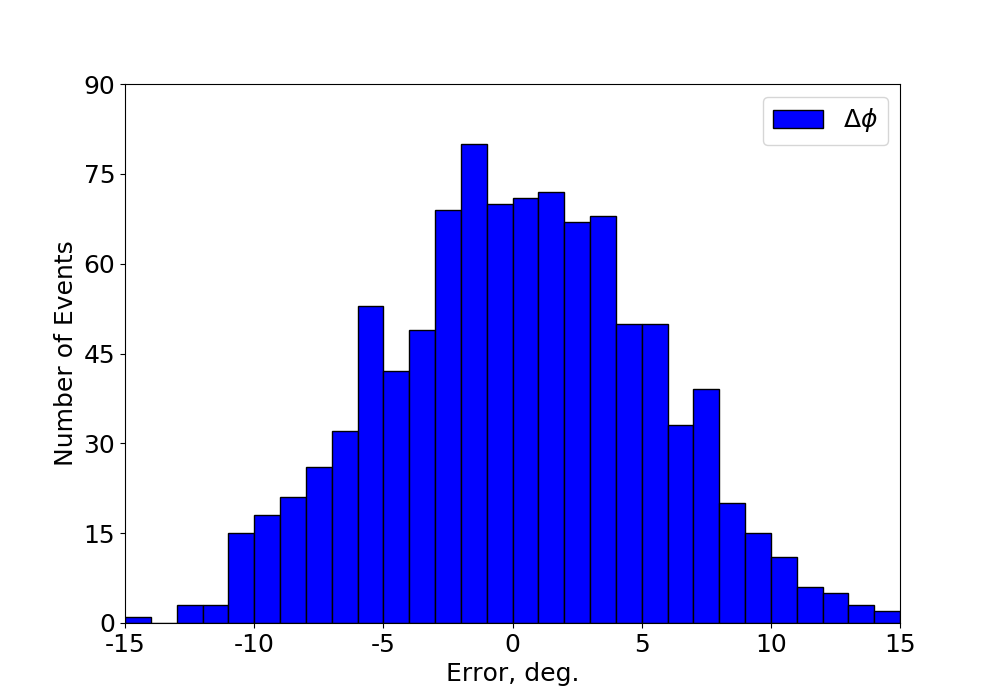}
\includegraphics[width=0.48\textwidth]{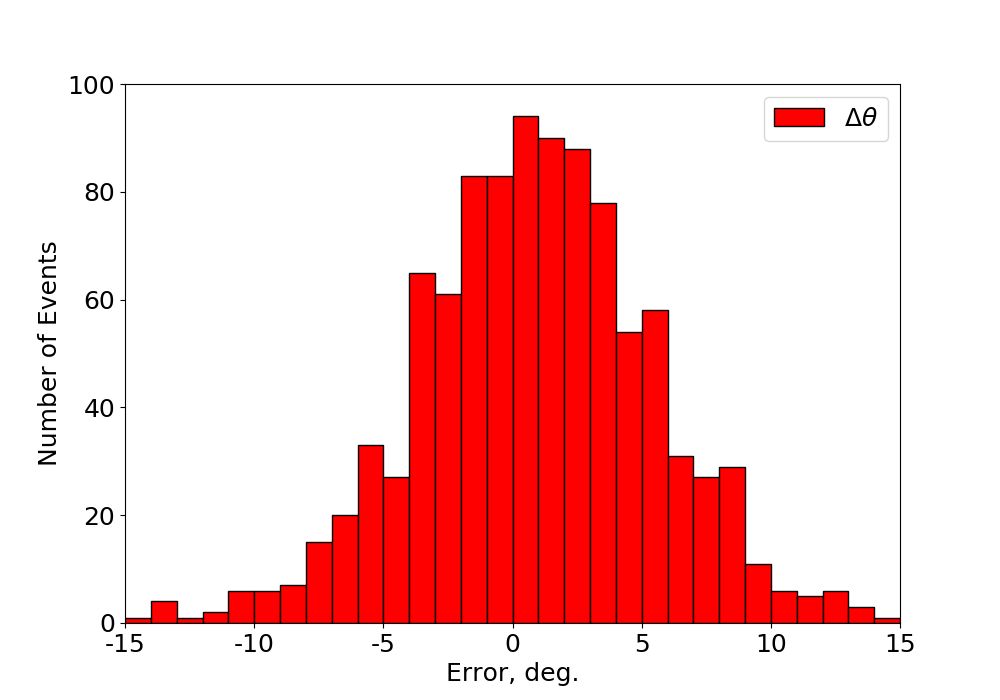}
\caption{Error Histograms for RDB for $\phi$ angle with the AV-prescriptions (left panel) and for $\theta$ angle for ZGP-prescription (right panel). $\Delta\phi = (0.2\pm5.3)^\circ$, $\Delta\theta = (0.9\pm4.7)^\circ$.}
\label{fig:EH_AV_ZGP}
\end{figure}

Histograms of the angle reconstruction for RDB events are presented on figure~\ref{fig:EH_AV_ZGP}. AV-prescription for $\Delta\phi$-histogram gives approximately zero bias, $\Delta\phi=(0.2\pm5.3)^\circ$ (here we use notation mean$\pm$std). By contrast, $\Delta\theta$-histogram constructed in accordance with ZGP-prescription, 
 has a noticeable bias, $\Delta\theta = (0.9\pm4.7)^\circ$. 
However, this bias depend strongly on data base parameters, so we can not fix this shifts in the $\theta$-reconstruction procedure.
Therefore, we will use the RMS errors as a measure of the LTA uncertainty, $\Delta\phi_{RMS} = 5.3^\circ$,  $\Delta\theta_{RMS} = 4.8^\circ$.  

\section{TUS events reconstruction}
\label{sec:TUSreco}

When reconstructing the TUS events, the following sequence of steps is performed:
\begin{itemize}
    \item Light curve reconstruction: $A_\mathrm{m}$, $T_\mathrm{m}$, FDHM, Asymmetry estimations by SkewGauss fit of the sum of all active signals.
    \item Track reconstruction by LTA: $U_y/U_x$, $X_\mathrm{m}$, $Y_\mathrm{m}$ by AV-prescription, $U$ by ZGP-prescription.
    \item Arrival direction: $\phi$ and $\theta$ from~(\ref{eq:Theta}), and corresponding direction on celestial sphere with 1st Equatorial coordinates. 
\end{itemize}

\begin{figure}[h]
    \centering
    \includegraphics[width=0.49\textwidth]{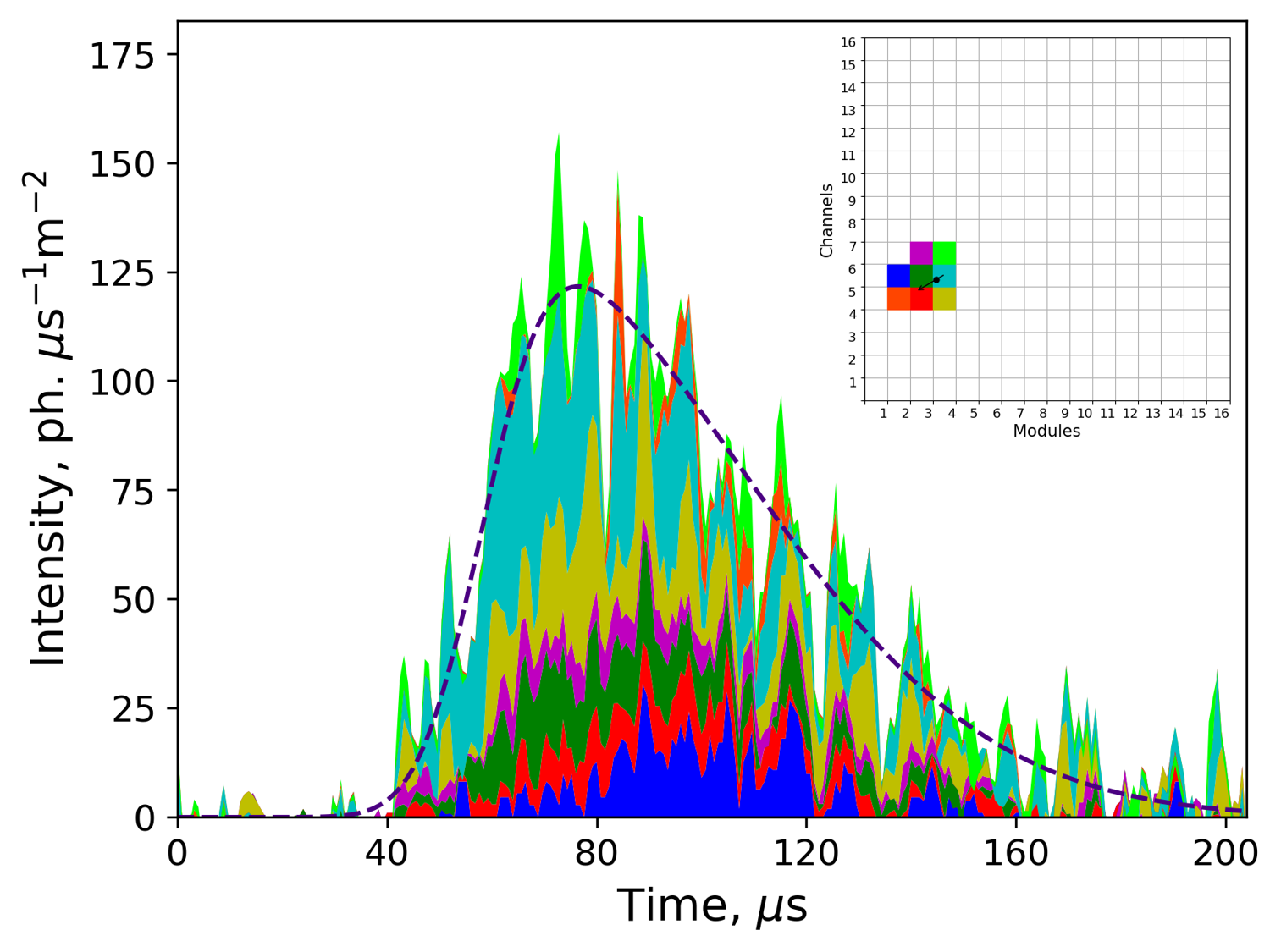}
    \includegraphics[width=0.49\textwidth]{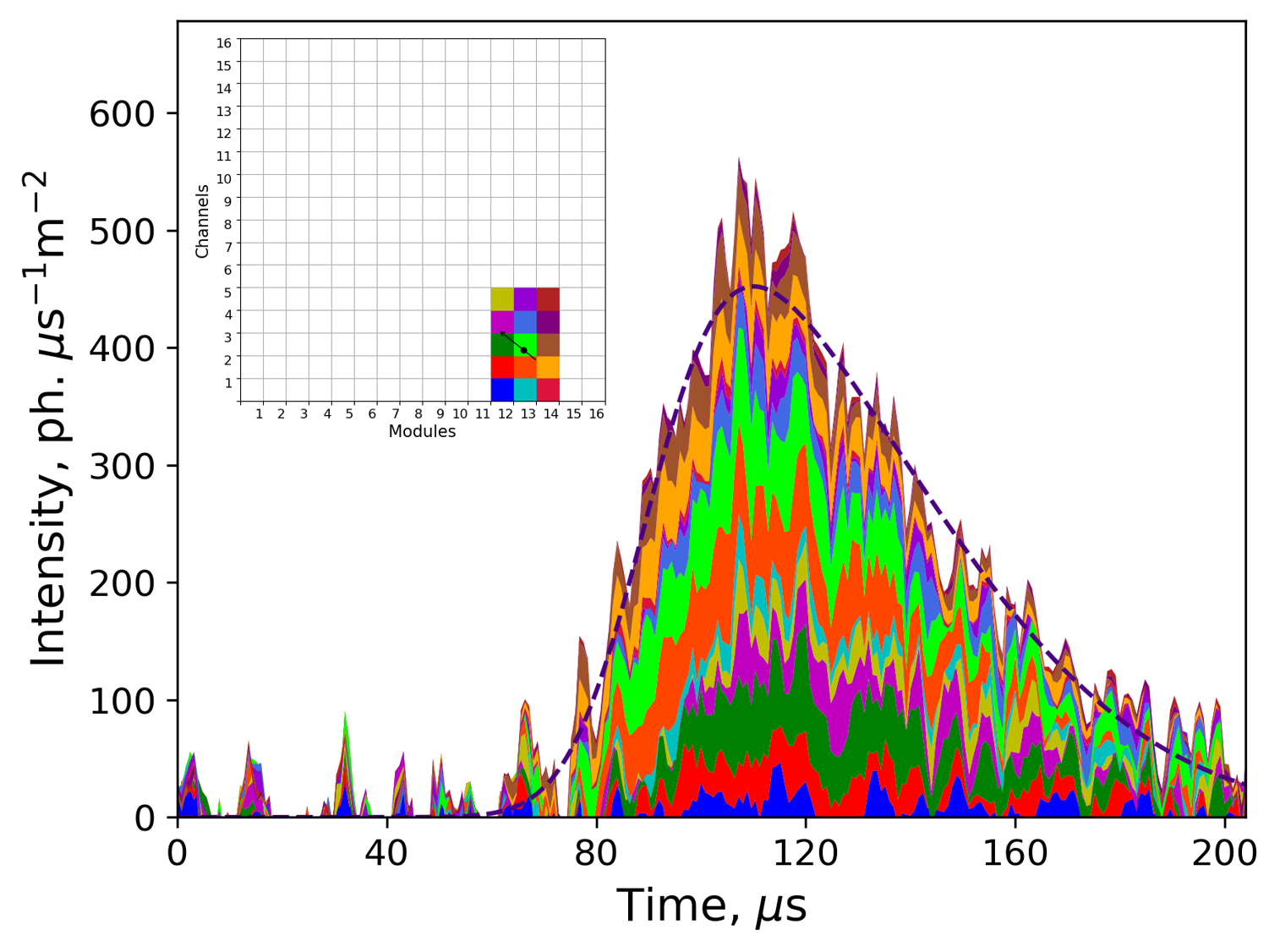}\\
    \includegraphics[width=0.49\textwidth]{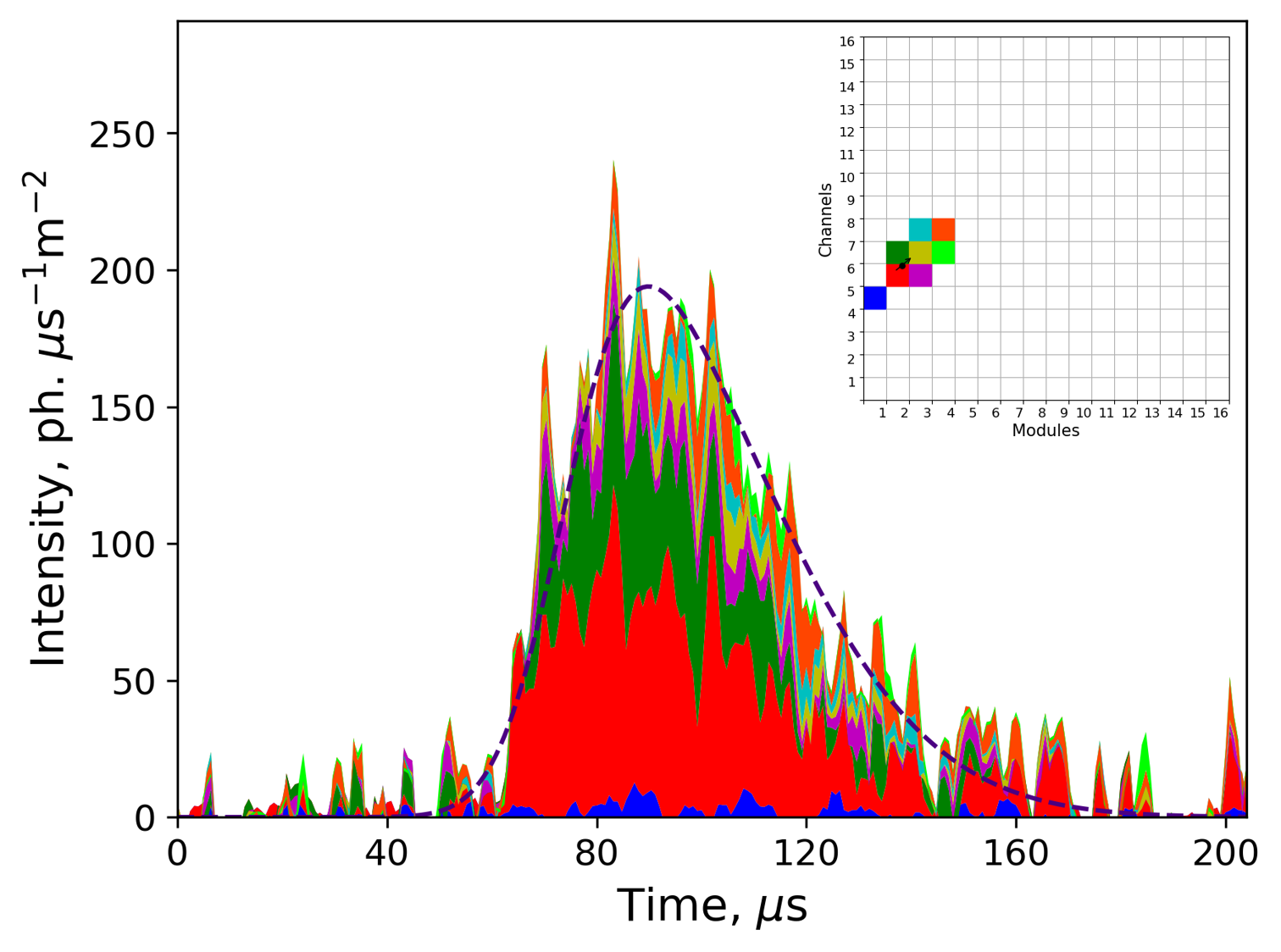}
    \includegraphics[width=0.49\textwidth]{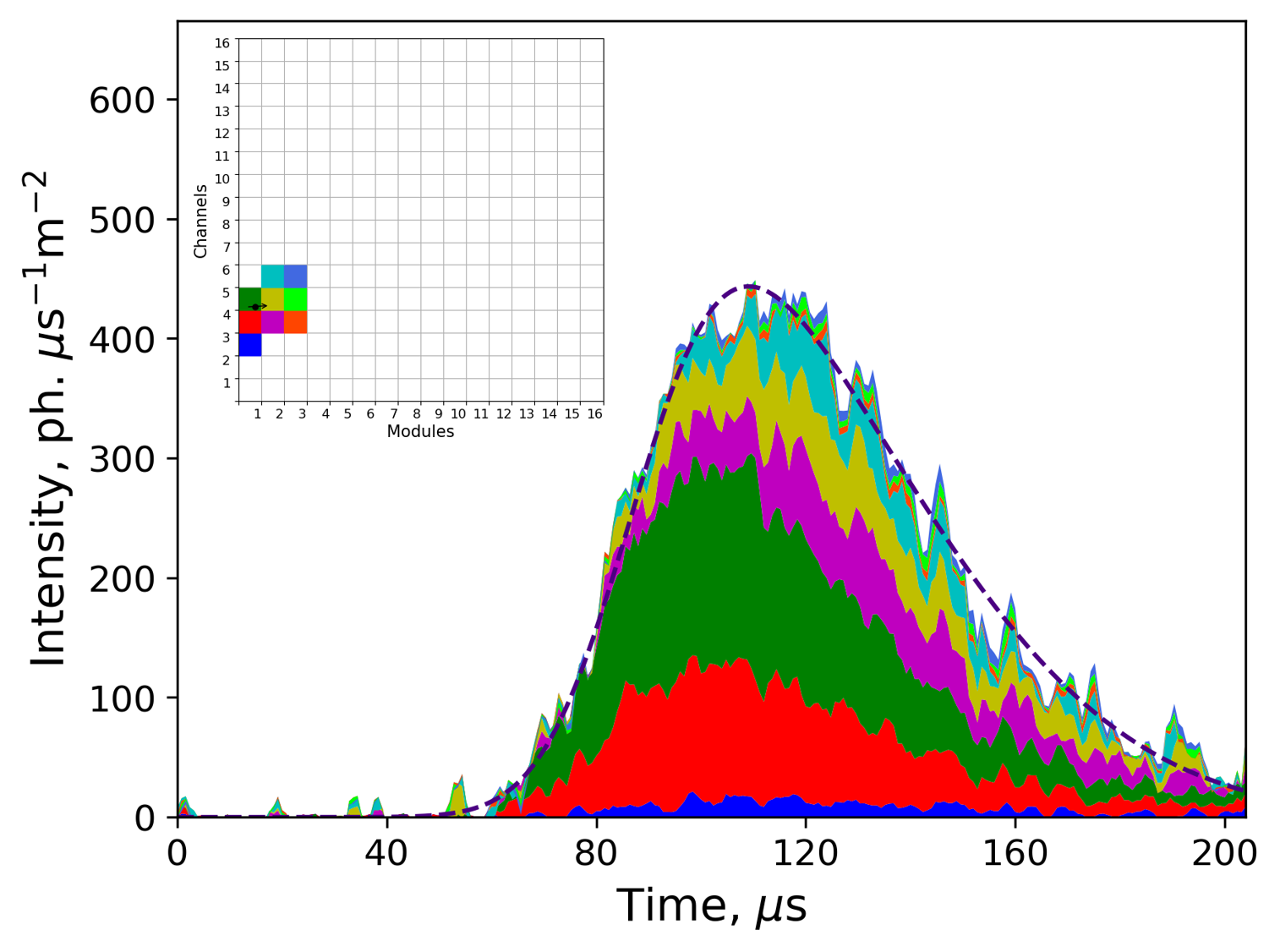}
    \caption{TUS events light curves: top -- TUS161031, TUS171010, bottom -- TUS171029a, TUS171029b.}
    \label{fig:events_LC}
\end{figure}

The results of the reconstruction of the light curves and the apparent kinematic parameters of the source movement for all EAS-like TUS events are presented in table~\ref{tab:RecoResults}. The light curves and corresponding pixel maps with reconstructed track (black arrow)  are shown\footnote{Light curve of the TUS170915 is presented on figure~\ref{fig:sample_LC}, and of the TUS161003 - in our previous paper~\cite{Minnesota}.} on figure~\ref{fig:events_LC}.

\begin{table}[!ht]
\caption{Light curve reconstruction (amplitude, duration and asymmetry coefficient) and apparent movement parameters.}
\begin{center}\small
\begin{tabular}{|l|*{6}{c|}}
\hline
Events & $A_\mathrm{max}$, ph.\,m$^{-2}$\,$\mu$s$^{-1}$ & FDHM, $\mu$s & Asymmetry & $\omega$, rad/s  & $V/c$ & $\Delta L$, km  \\
\hline
TUS161003 & 205 & 49.4 & 1.56 & 210 &  0.34  &  10.0 \\
\hline
TUS161031 & 121 & 61.8 & 2.18 & 122 & 0.20 & 7.4 \\
\hline
TUS170915 & 442 & 72.0 & 1.69 & 173 & 0.28  &  12.1 \\
\hline
TUS171010 & 452 & 62.8 & 1.82 & 179 & 0.28  &  10.6 \\
\hline
TUS171029a & 193 & 46.6 & 1.59 & 112 & 0.18  & 4.9  \\
\hline
TUS171029b & 443 & 63.8 & 1.61 & 81 & 0.13  &  4.9 \\
\hline
\end{tabular}
\end{center}
\label{tab:RecoResults}
\end{table}

Part of the luminescence cannot be recognized due to rather large PSF, so the reconstruction gives the deformed shape of the light curve. The estimates of the $A_\mathrm{m}$, FDHM, and Asymmetry have noticeable errors, e.g. $A_\mathrm{m}$ is lower estimate of the amplitude. A significant contribution to the uncertainty of these quantities is also made by the inaccuracy of the absolute sensitivity calibration (see below about ``red'' pixels).

The angular velocity $\omega=U/f$ of the line of sight to the source characterizes its movement in the picture plane only and does not depend on the distance to the source. On the contrary, the apparent linear velocity $V=\omega R$ and track length $\Delta L = V\cdot$ 2FDHM (both projected onto the picture plane) were obtained under the assumption that the distance to the source coincides with the height of the satellite’s orbit (varies from 475 to 495~km). One can see that all events are relativistic. 

When reconstructing the TUS events in comparison with simulated ones, two conditions,
1) strong irregularity in the distribution of channel sensitivities, and 2) asymmetry of PSF (non-isotropic and non-uniform) should be taken into account. Both primarily affect the estimates of the velocity components of movement along the track.

The existence of channels with low sensitivity leads to
skipping signals in some cases, and to not reliable fitting in others. In addition, the calibration of less sensitive channels has the largest uncertainty (for several channels, the sensitivity coefficients are known with an accuracy of hundreds of percent). We will call such weakly sensitive channels ``red''. In those cases when the selection procedure finds ``red'' pixels, we plot reconstruction curves for a set of sensitivity values and estimate the variability of the AV- and ZGP-values (see below).  

The asymmetry of the PSF can lead to a significant systematic bias in the speed estimation of the image propagation along the track. For example, an off-axis coma-type aberration shifts the image centroid in a radial direction relative to the position of the non-aberration track point. This stretches the track and leads to an overestimation of speed by the LTA method.
It is very difficult to quantify this effect. Reconstruction of simulated events with a convolution of the coma and Gauss distributions (with the TUS mirror parameters) as PSF showed that the systematic bias in $\theta$ can vary from 0 to +3.5$^\circ$ depending on the location of the track on the pixel map. To partially reduce this bias, the signals fit should be performed by an asymmetric function that selects the main part of the signal not near the centroid, but close to the track point. We use the SkewGauss function to fit the EAS-like TUS event signals (with a limited range of shape and scale parameter).
So besides the methodical inaccuracy, an additional event specific  contribution to the uncertainty of the reconstructed angle should be taking into account, we name it the \textit{measurement error}.

\begin{table}[!ht]
\caption{Arrival direction in local ($\phi,\theta$) and Equatorial coordinate systems ($\alpha,\delta$). Number of weakly sensitive channels ('reds'), measurement error for $\phi$ and special feature of registration are presented in the last column (see the text for details).}
\begin{center}\small
\begin{tabular}{|l|*{5}{c|}}
\hline
 Events & $\phi$, $^\circ$ & $\theta$, $^\circ$ & $\alpha$, $^\circ$ & $\delta$, $^\circ$ & Comments  \\
\hline
TUS161003 & 144 & 38 & 308 & 65 & \\
\hline
TUS161031 & 166  & 26 & 343 & 83 & 2 reds, $\Delta\phi_\mathrm{mes}=2^\circ$
\\
\hline
TUS170915 & 86 & 28 & 311 & 33 & $\Delta\phi_\mathrm{mes}=1^\circ$\\
\hline
TUS171010 & 231 & 28 & 41 & 49 & \\
\hline
TUS171029a & 333 & 17 & 35 & 19 & 1 red, $\Delta\phi_\mathrm{mes}<1^\circ$\\
\hline
TUS171029b & 16 & 11 & 33 & 55 & boundary event\\
\hline
\end{tabular}
\end{center}
\label{tab:RecoDirection}
\end{table}

In the following, the reconstruction of the arrival directions was made under the assumption that the total speed of the source is equal to the speed of light.
The results are presented in table~\ref{tab:RecoDirection}, 
where the direction is expressed both as the ($\phi$, $\theta$) angles of the local coordinate system, and in the Equatorial coordinate system (right ascension $\alpha$ and declination $\delta$). 
The directions of arrival of the sources of the registered events are not concentrated near any one direction, but are distributed over a significant part of the celestial sphere.

\begin{figure}
    \centering
    \includegraphics[width=0.45\textwidth]{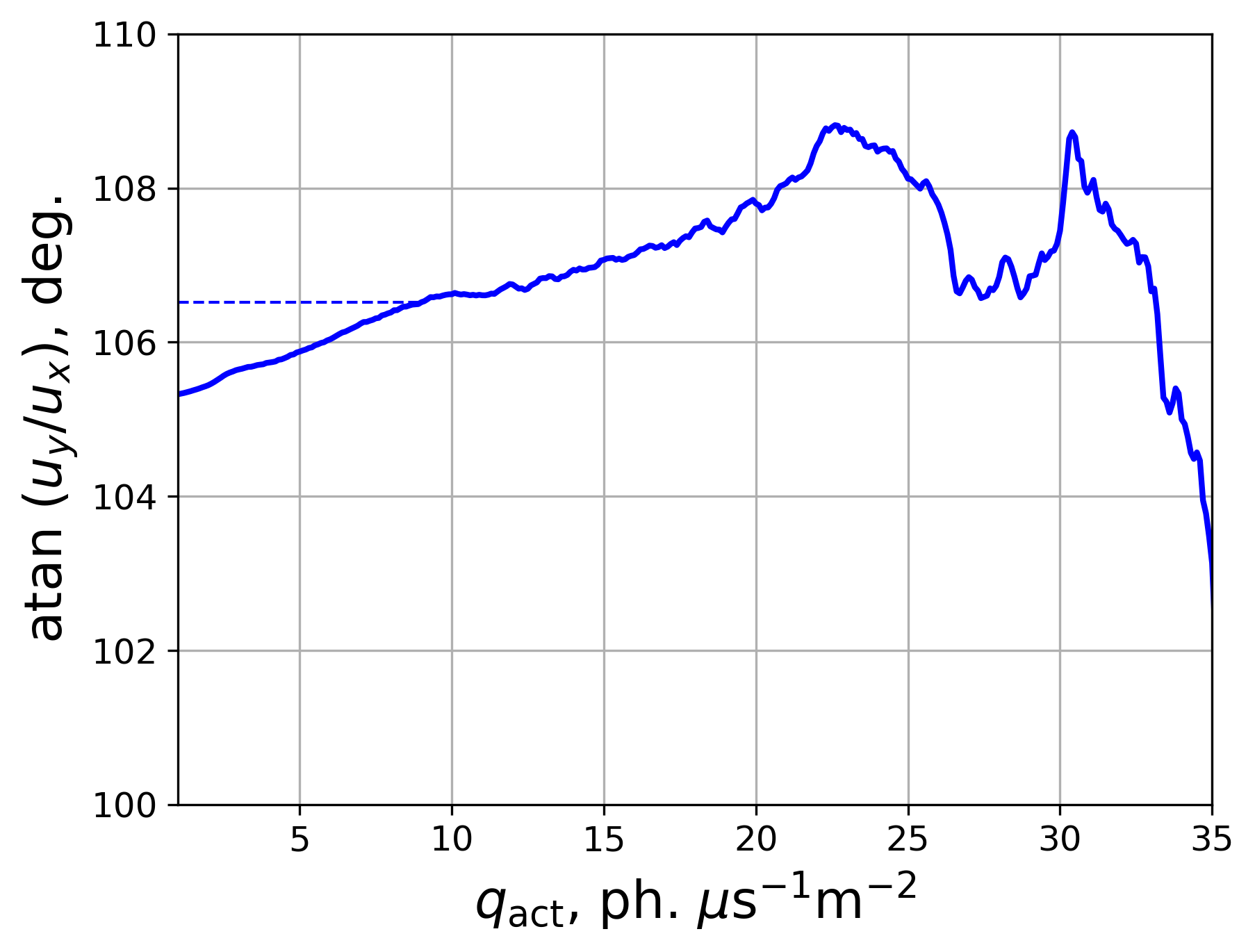}
    \includegraphics[width=0.45\textwidth]{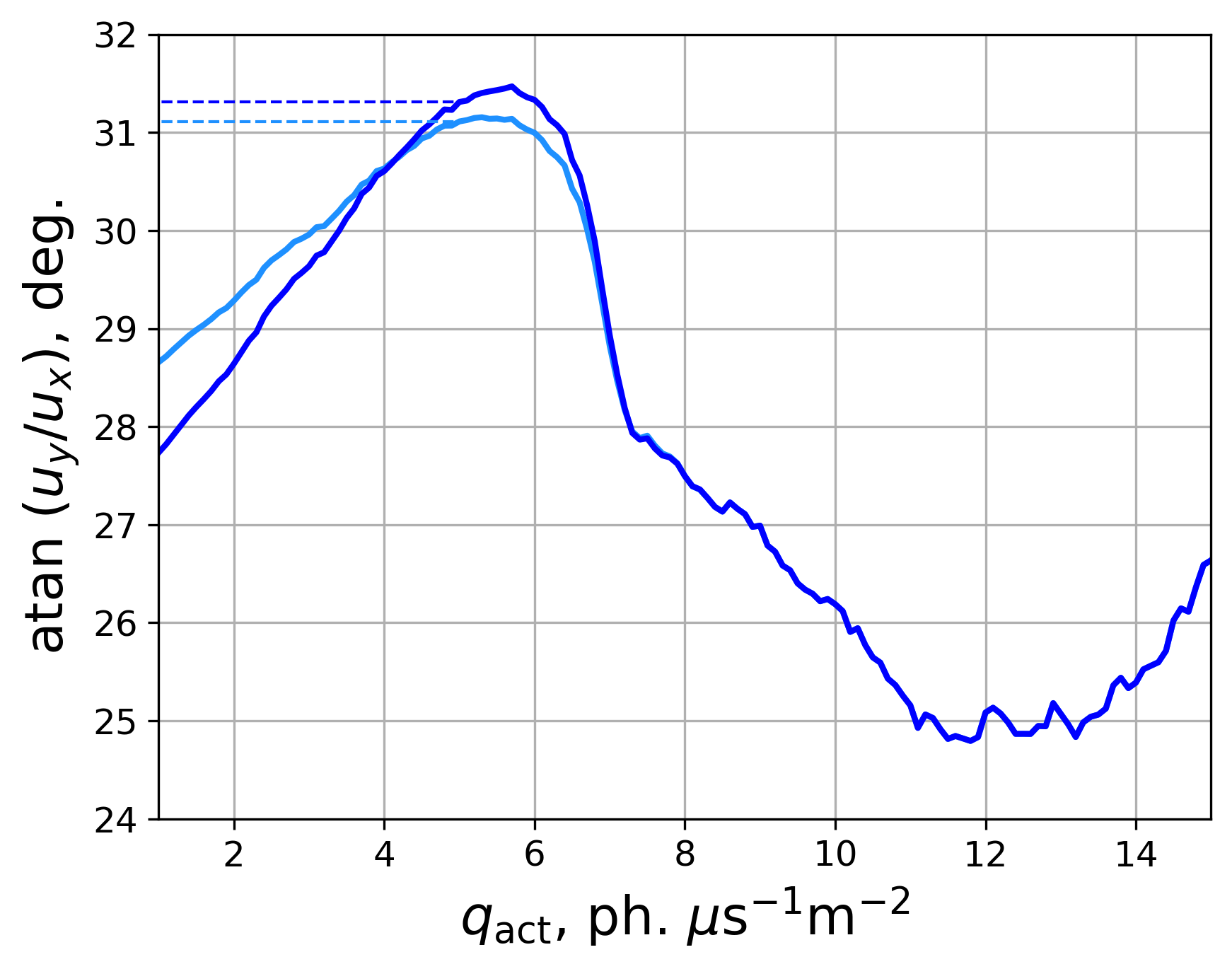}
    \caption{AV-region determination: left panel - for the event TUS170915, right panel - for the event TUS161031 (two variants, see text).}
    \label{fig:AV-region}
\end{figure}

The TUS170915 is an example of an event whose direction of arrival can be determined with a high degree of reliability. 12~pixels were selected to build the reconstruction curves in this case. The shape of the $\theta$-curve is similar to the typical one with definite ZGP. However, in contrast to the simulated $\phi$-curves, the real one (figure~\ref{fig:AV-region}, left panel) has a small gradient at the smallest threshold $q_\mathrm{act}$, which makes it difficult to determine the AV~region. In fact, the presence of such a gradient is typical of EAS-like TUS events and can be explained by the existence of significant asymmetry of pixel signals caused by the contribution of coma aberration due to~PSF. For the TUS170915, this ``deformation'' of the $\phi$-curve is insignificant, and the 'truth' AV~region can be detected near $q_\mathrm{act} = 10$ (see the dashed line in figure~\ref{fig:AV-region}). So for the TUS170915, the measurement error is estimated as $\Delta\phi_\mathrm{mes} = 1^\circ$.

For the event TUS161031, only six selected channels have a reliably sensitivity coefficients, two more are ``red''. Accordingly, a set of reconstruction curves was used with different values of the sensitivity for the red channels. Right panel of figure~\ref{fig:AV-region} shows $\phi$-curves for two limit values ($\theta$-curves have a typical shape and vary slightly). The measurement error is estimated as $\Delta\phi_\mathrm{mes} = 2^\circ$.

A weakly sensitive channel is also found in the event TUS171029a. However, in this case, the reconstruction curves practically do not change when the sensitivity coefficient varies over a wide range for this red channel. Therefore, the measurement error is not more than~$1^\circ$. 

Particular difficulty arises during the reconstruction of the event TUS171029b. It is located at the edge of the detector FOV and the analysis of the signals shows that part of the track lies outside, and therefore has not been registered (see the inset on the bottom-right panel in figure~\ref{fig:events_LC}). The presence of a signal outside the FOV can ``tilt'' the track during reconstruction. In addition, in such cases of ``boundary events'', the LTA underestimates the determination of a speed of movement along the track. Thus, a measurement error is present here for both the azimuth and zenith angles, however, it is currently not possible to evaluate it (so azimuth reconstructions in this case are qualitative rather than quantitative).

Finally, in the two remaining events, TUS171010 and TUS161003, the above problems are not observed: the large number of selected pixels with relatively high sensitivity allows us to estimate the angles without additional error. 
In~\cite{Minnesota}, the zenith angle of the TUS161003 was estimated $6^\circ$ more, a consequences of using a less reliable version of LTA with a relative rather than an absolute threshold $q_\mathrm{act}$. A more thorough analysis showed that such an LTA option in some cases can lead to large systematic shifts in the $\theta$-estimation.

\section{Discussion}
\label{sec:Disc}

\begin{figure}
    \centering
    \includegraphics[width=0.65\textwidth]{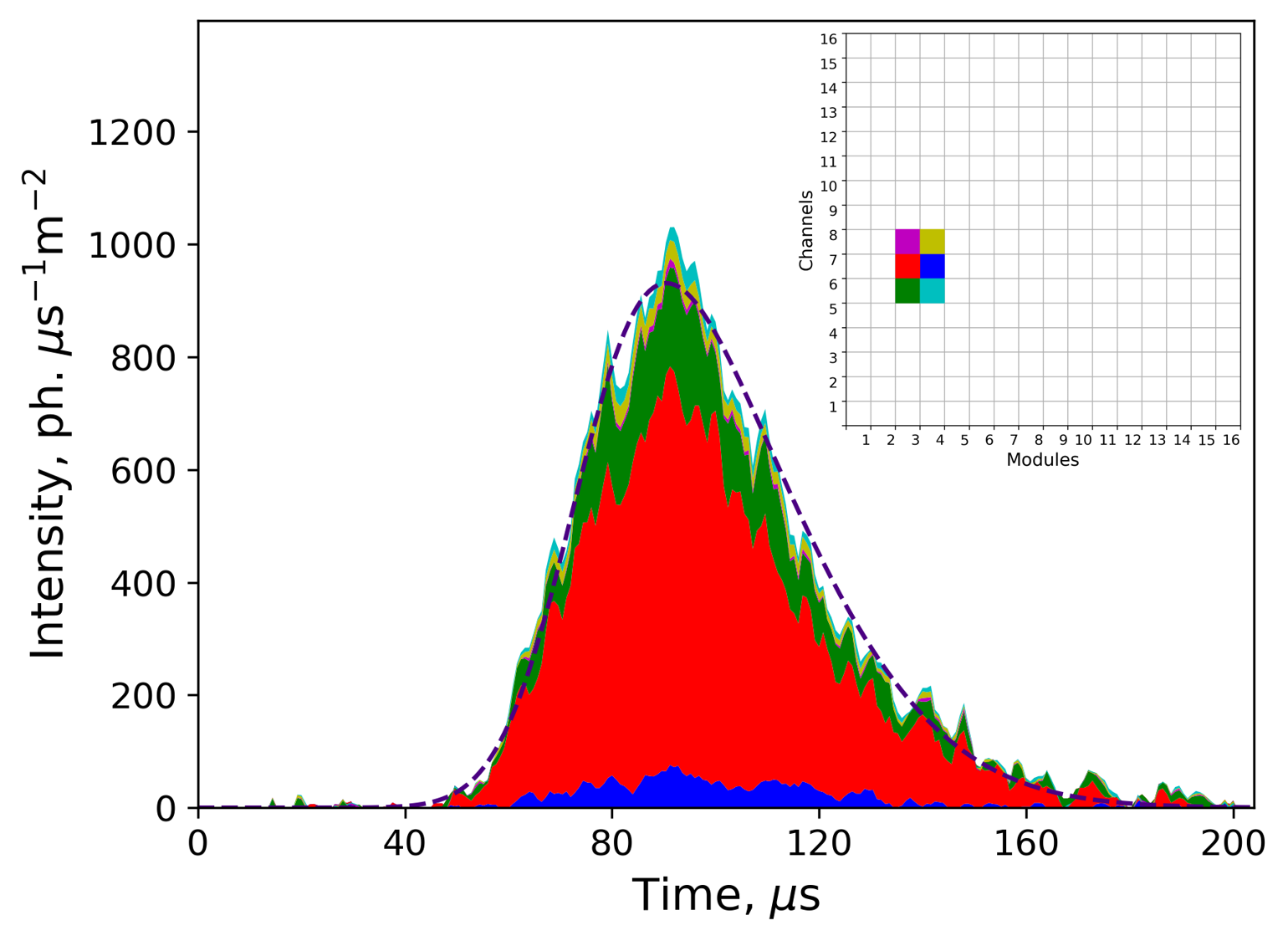}
    \caption{One of the TUS events with expected EAS light curve but without peak movement.}
    \label{fig:no_peak_mov_LC}
\end{figure}

We emphasize that the choice of these six events as representatives of the EAS-like class was determined by two features: the characteristic shape and duration of the light curve and the presence of a noticeable time displacement of the channel signal peaks. This last cut can be interpreted as the movement of a relativistic radiation source. The LTA has been tested only on events with such a movement.
A number of recorded events partially suited to these properties were not analyzed in this work.
For example, the figure~\ref{fig:no_peak_mov_LC} shows the light curve of another event registered 27.06.2017 above the USA (32.89S, 100.99W), for which, despite the presence of 6 hit pixels, it is not possible to establish the source movement: LTA reconstruction leads to $\theta<1^\circ$, and the image itself is most likely caused by the PSF effect.

On the other hand, a lot of events with EAS-like light curves were rejected due to the small number of hit pixels since the reconstruction is unreliable in this case.

The formulas presented in section~\ref{sec:Intro} for the non-relativistic, relativistic, and ultra-relativistic cases can be unified as
\begin{equation}
v/c = \frac{\sin\chi_0}{\sin(\chi-\chi_0)},\quad \tan(\chi_0)=V/c,
\end{equation}
For a fixed value of the apparent linear velocity $V$, different geometries of the track correspond to a different speeds of a source. There are no tracks with $\chi<2\chi_0$.

A minimum speed of the moving object, $v_{\min} = c \sin\chi_0$, corresponds to up-going track at an angle of $\pi/2 + \chi_0$ with respect to the line of sight.
For typical value of $V/c=0.25$ from the table~\ref{tab:RecoResults}, we have $\chi_0\approx14^\circ$ and $v_{\min}/c = 0.24$.

For all other arrival directions, two possible $v$ correspond to the observed apparent velocity~$V$. It is important to emphasize that for all 6 estimations of the $V$ and any variant of the arrival direction the motion of the source is  relativistic (as $v\ge v_{\min}$). 

It is rather difficult to assume the presence of a relativistic source of UV radiation, with speed different from the speed of light, for example, in the range from 0.2 to $0.9c$.
In the ultra-relativistic limit, the movement can be either down-going, $\chi\to2\chi_0$, or up-going vertically along the line of sight, $\chi\to\pi$. The second scenario, however, cannot be realized in our events, since the observed luminescence track length would correspond to a very large track in the atmosphere (more than 100~km), the region where UV radiation cannot be deliberately produced. So the most probable kinematics of EAS-like TUS events corresponds to scenario with $\chi=2\chi_0$ which we've used in the LTA reconstruction.

\begin{figure}
    \centering
    \includegraphics[width=0.95\textwidth]{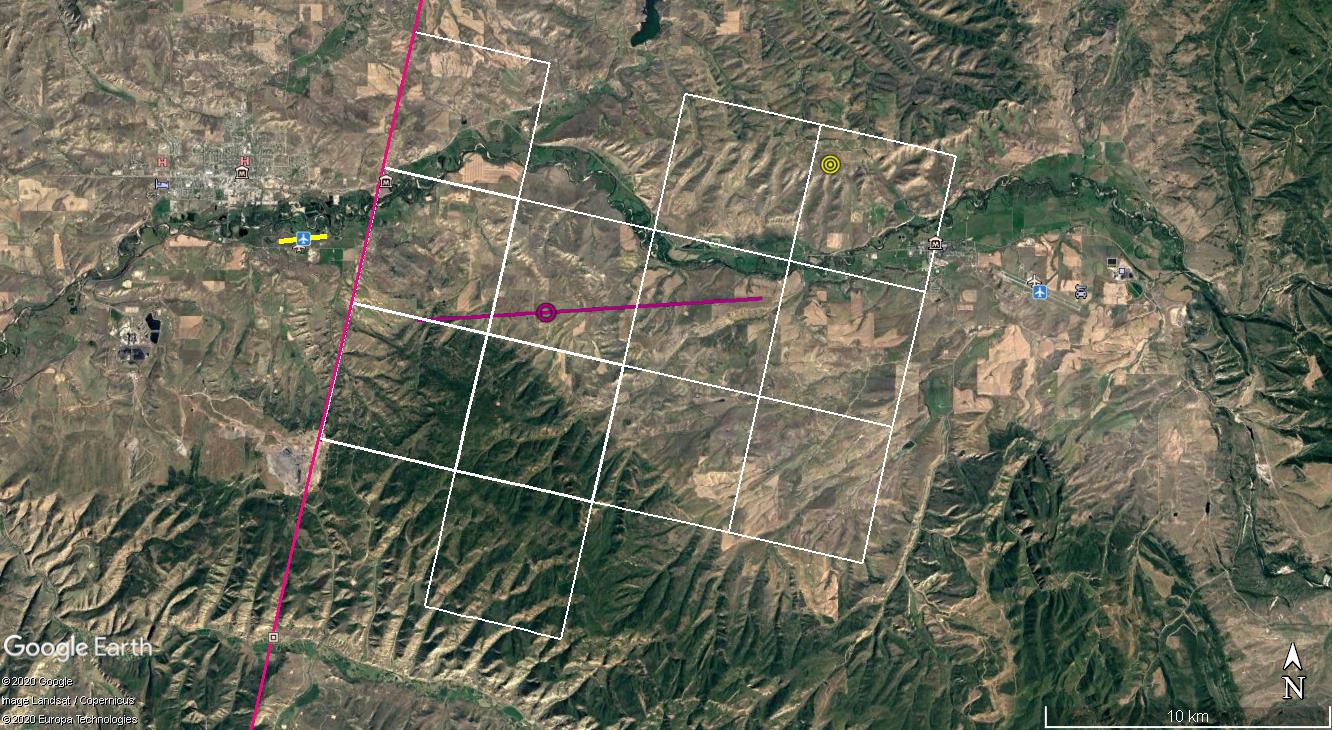}
    \caption{Google Earth map for the event TUS170915. The yellow line corresponds to the runway of the Craig Moffat County airport. The reconstructed track presented by magenta line with a circle ('maximum point').}
    \label{fig:TUS170915_GE}
\end{figure}

Using the Google Maps resource\footnote{\url{https://www.google.com/earth/}}, an analysis was made of objects localized at the track location of each of the 6 events. It turned out that in all cases there is one or more airports in the immediate vicinity of the track. 
In particular, as shown in the figure~\ref{fig:TUS170915_GE}, the TUS170915 track (the dark magenta line with a circle) is a continuation of the runway of the \textit{Craig Moffat County Airport} within the accuracy of the reconstruction. The white squares in the figure show the fields of view of 12 hit pixels ($\sim5\times5$~km$^2$), the magenta circle indicates the position of the maximum.

In this regard, an assumption was made about the anthropogenic origin of the source of all the events considered in this article. More precisely, a hypothesis was put forward on the correlation in the position and direction of runways and the observed track.
\textit{Airport Runway hypothesis} means that the track of the event coincides with the continuation of the airport runway which involves two conditions: 1) small difference $\Delta\phi$ in track and runway directions, and 2) their close location.

The accuracy of the angle between the projection of the direction of the track onto the Earth’s surface and the direction of the runway $\Delta\phi$ corresponds to the accuracy of the reconstruction of the azimuth angle $\phi$.
In addition to the methodical component ($\sim\pm5^\circ$), $\Delta\phi$ has measurement error varying from event to event.
The main uncertainty in the position of the track is due to the inaccuracy of the time stamp of the registration of the event. In fact, the trigger time stamp was transmitted with an accuracy of 1 second, which corresponds to shifts of the detector FOV along the direction of motion by about 8 km (cf. with 5~km size of pixel field of view). In addition, sometimes de-synchronization occurred during the TUS operation, which led to uncontrollable time shifts $\Delta t$ of 1-3 seconds.

For each EAS-like event we are looking for the closest airport  and estimate $\Delta\phi$ angle between track and runway directions and probable time shifts $\Delta t$. 
The tracks of five of the six EAS-like events are similar in location and direction to the runway:
\begin{itemize}
\item Event TUS161031: \textit{Sparrevohn AFC Airport with LRRS}, $\Delta\phi= 10^\circ$, $\Delta t<0.5$~s (track from or above runway).
    
\item Event TUS170915: \textit{Craig Moffat County Airport}, $\Delta\phi \approx0^\circ$, $\Delta t<0.5$~s (from runway).

\item Event TUS171010, two candidates: 1) \textit{Cherry Point Mcas Airport}, $\Delta\phi < 5^\circ$, $\Delta t\approx +2$~s (from runway) or 2) \textit{Coastal Carolina Regional Airport}, $\Delta\phi = 15^\circ$, $\Delta t\approx -1$~s (from or above runway).  

\item Event TUS171029a: \textit{Winslow-Lindbergh Regional Airport}, $\Delta\phi \approx 25^\circ$ (from runway), $\Delta t\approx -0.5$~s (low accuracy of track reconstruction). 

\item Event TUS171029b: \textit{Tin City Airport (with LRRS)}, $\Delta\phi = 20^\circ$, $\Delta t\approx0$~s (above runway). This is ``boundary event'' with a big part of the signal outside the TUS FOV.
\end{itemize}
We can conclude that in the first three cases, the hypothesis is confirmed within the accuracy of measurements. In the last two events, the additional error in $\Delta\phi$ can be explained by a peculiar properties of the reconstruction. 

The only event that clearly does not correspond to the hypothesis is  one which detailed analysis was carried out in~\cite{Minnesota}:
\begin{itemize}
\item Event TUS161003: \textit{Christison Airport}, $\Delta\phi \approx 45^\circ$, $\Delta t\approx -1$~s (to runway).  
\end{itemize}

In the vicinity of the event there are a large number of local airports, however, the orientation of all their runways is very different from the direction of the track. It is interesting to note that the only nearby sufficiently large \textit{Rochester International Airport} has the ``correct'' orientation ($\Delta\phi<5^\circ$), but it is shifted (along the detector movement) by 40~km.

There are several possible explanations for this direction--location correlation. On the one hand, the source of linearly aligned UV light can be aircraft warning lights  systems located along the direction of approach to a runway. 
In addition, mounted on aircraft landing lights illuminate the terrain and runway ahead during takeoff and landing. This illumination is bright white, forward and downward facing lights on the front of an aircraft (the landing lights of large aircraft can easily be seen by other aircraft over 100 miles away). 

Unfortunately, in any of these variants it is rather difficult to explain the characteristic spatio-temporal pattern of events recorded by the TUS: a light curve with a duration of about 100~$\mu$s and a signal peak displacement with an apparent speed of several tens of percent of the speed of light.

It is well known that transient UV flashes can also be observed when an electromagnetic pulse (EMP) penetrates into the lower layers of the ionosphere, \cite{Wilson_1924}. In particular, the appearance of the so-called elves, large-scale ring-shaped optical flares at an altitude of 80-90 km, with a characteristic glow time of about 1 ms, is explained by an EMP generated by lightning discharge, \cite{doi:10.1029/95GL03816},  \cite{doi:10.1029/2009JA014775}.
During its operation, the TUS detector registered 25 elves (including the so-called double elves), \cite{garipov2019remote245355787}. 
For this type of events, the characteristic duration of the signal in a channel was also several tens of microseconds, but its amplitude was more than an order of magnitude higher, and the whole image of the elve occupied most of the entire FOV.

Several different types of radio transmitting stations are located near airports\footnote{\url{https://en.wikipedia.org/wiki/Instrument_landing_system}}. VHF omnidirectional range (VOR) is used to determine the azimuth of the direction of movement of the aircraft (often combined with Distance Measuring Equipment, DME). VOR is a phased antenna array to send a highly directional signal that rotates 30 times a second.
For each of the EAS-like TUS events, we have identified the nearest VOR station using the resource \url{https://www.airnav.com}. For example, in  figure~\ref{fig:TUS170915_GE}, Hayden VOR/DME station (40.5201$^\circ$N, 107.3049$^\circ$W) is marked with a yellow circle, so the TUS170915 track  is located exactly between (with an accuracy of an acceptable offset within 0.5 s) the runway and the VOR. 

\begin{figure}
    \centering
    \includegraphics[width=0.95\textwidth]{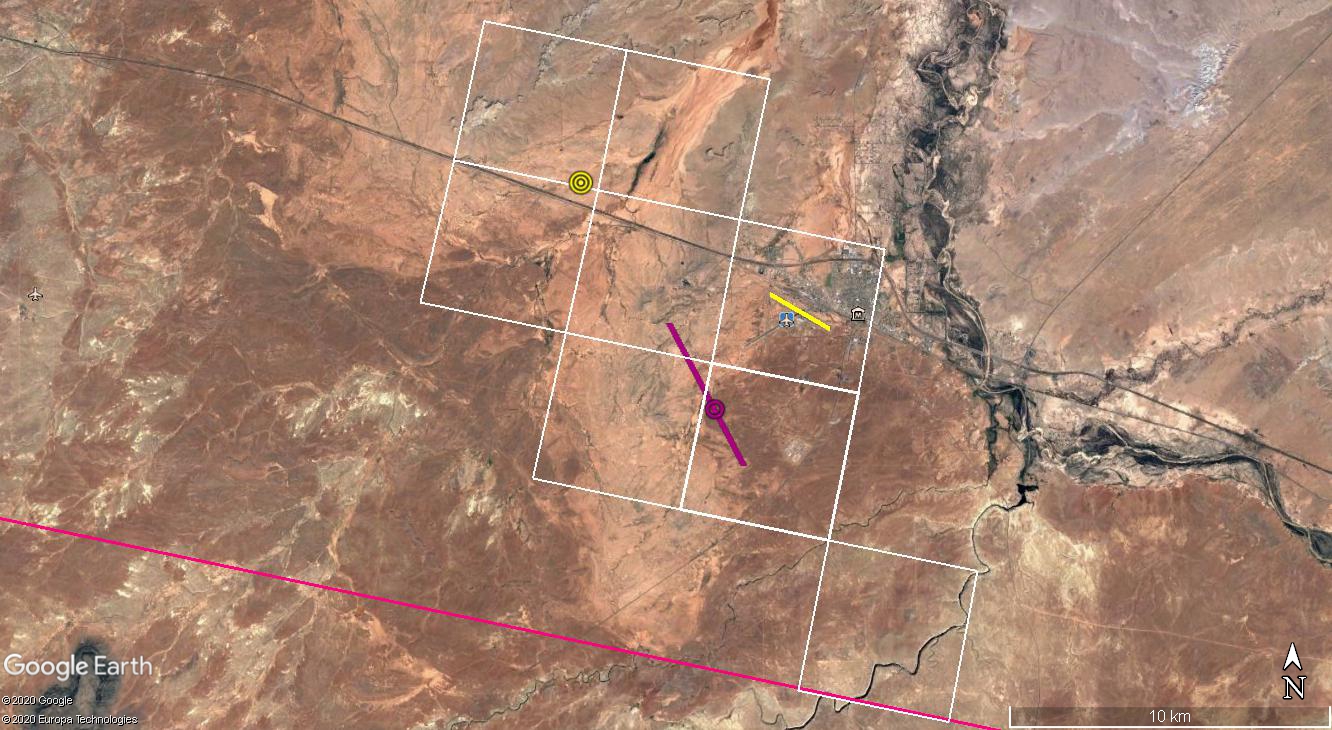}
    \caption{Google Earth map for the event TUS17102a. The yellow line corresponds to the runway of the \textit{Winslow-Lindberg Airport}, and yellow circle to the  Winslow VORTAC station.}
    \label{fig:TUS171029a_GE}
\end{figure}

Interestingly, despite the directions of the TUS17102a track and the runway do not coincide, Winslow VORTAC\footnote{VORTAC is a collocated VOR and Tactical Air Navigation station.} station (35.0616$^\circ$N, 110.7950$^\circ$W) is located at the intersection of these two directions, see figure~\ref{fig:TUS171029a_GE}.
For all EAS-like TUS events, a VOR station was found in the immediate vicinity of the event maximum. Specifically for the TUS171010 event, New Bern VOR/DME (35.0731$^\circ$N, 77.0451$^\circ$W) is located right next to the \textit{Coastal Carolina Regional Airport}. 
The only exception is again the TUS161003 event: here  VOR/DME is located on the extension of the \textit{Rochester International Airport} runway, i.e. more than 30~km from the reconstructed track.

However, the hypothesis of the appearance of a EAS-like events by means of ``highlighting'' the lower layers of the ionosphere with a VOR radio signal has many shortcomings. The main energy of the lightning EMP which generates elve is carried in the VLF (3-30 kHz) range, while the radio signal generated by the VOR station has a frequency in the 108-118 MHz range, and the VOR transmitter power is much weaker to the lightning. 
In addition, in this case the apparent speed of movement when observed from orbit must exceed the speed of light.

\section{Conclusions}
The orbital detector TUS, due to the use of a large-area mirror-concentrator  and a highly sensitive photodetector, is able to register short-term UV signals (flashes) that are inaccessible to other devices. In this work, special attention was paid to the class of so-called EAS-like events with a characteristic light curve duration of about 100~$\mu$s and a noticeable
spatial displacement of the peak of the signal (``track'').
A total of six such events were selected. 
Reconstruction of such events is complicated by a short track length, an asymmetry of the image, and an uncertainty in the sensitivity distribution of the TUS channels.
The new method of reconstruction of the kinematic parameters of such events was developed and made it possible to determine the direction of arrival of their source - it corresponds to downward relativistic movement with zenith angles from~$10^\circ$ to~$40^\circ$.

Despite their name, it is difficult to associate such events with the EAS fluorescence because of significant signal intensity (it should correspond to ZeV primary). All six events were located over the mainland of the United States. Moreover, the reconstructed tracks revealed a correlation with the runways of the airports. 
This allowed us to propose several possible hypothesis for the anthropogenic origin of the sources of these phenomena, including aircraft landing lights and the effect of radio pulses from VOR stations on the ionosphere, each of which faces difficulties in interpreting all the revealed patterns.
Detailed analysis of these and other possible hypotheses of the origin of the EAS-like events will be carried out in another paper.

The method developed in this work provides the reliable determination of kinematic parameters of the track-like events, registered with low spatial resolution. 
For the final solution of the riddle of the origin of events of this type, it is extremely important to increase their statistics. 
In this, the analysis of data from the new orbital detector Mini-EUSO~\cite{mini-Capel-2018} can play a key role.
In the case of anthropogenic origin of EAS-like events, the determination of their characteristics will make it possible to exclude them from further analysis in next-generation space missions K-EUSO~\cite{K-EUSO2017} and POEMMA~\cite{POEMMA} (or even implement suitable selection criteria into trigger systems).

\acknowledgments
The authors are grateful to Mikhail Zotov for his very important work in the search and classification of TUS events, and to Pavel Klimov and Boris Khrenov for a fruitful discussion of the results.

Oliver Isac Ruiz Hernandez is grateful to Oscar Martínez Bravo by his guidance and continuously supervision on his process to become a researcher.

The work was done with partial financial support from the State Space Corporation ROSCOSMOS, M.V. Lomonosov Moscow State University through its “Prospects for Development” program (“Perspektivnye Napravleniya Razvitiya”).
Also with partial financial support of the Vicerrectoría de Investigación y Estudios de Posgrado (VIEP-BUAP).

\bibliographystyle{JHEP}
\bibliography{references}

\end{document}